\documentclass[twocolumn, times, tighten]{aastex631}

\usepackage{CJK}

\newcommand{\Msun}{\ensuremath{{M_{\odot}}}}
\newcommand{\Lsun}{\ensuremath{{L_{\odot}}}}
\newcommand{\logM}{\ensuremath{{M_{\star}/M_{\odot}}}}

\providecommand{\e}[1]{\ensuremath{\times 10^{#1}}}

\shorttitle{SMILES Initial Data Release}
\shortauthors{Alberts et al.}

\begin{document}

\title{SMILES Initial Data Release: Unveiling the Obscured Universe with MIRI Multi-band Imaging}

\correspondingauthor{Stacey Alberts}
\email{salberts@arizona.edu}

\author[0000-0002-8909-8782]{Stacey Alberts}
\affiliation{Steward Observatory, University of Arizona,
933 North Cherry Avenue, Tucson, AZ 85719, USA}


\author[0000-0002-6221-1829]{Jianwei Lyu (\begin{CJK}{UTF8}{gbsn}吕建伟\end{CJK})}
\affiliation{Steward Observatory, University of Arizona,
933 North Cherry Avenue, Tucson, AZ 85719, USA}

\author[0000-0003-4702-7561]{Irene Shivaei} \affiliation{Centro de Astrobiolog\'ia (CAB), CSIC-INTA, Ctra. de Ajalvir km 4, Torrej\'on de Ardoz, E-28850, Madrid, Spain}

\author[0000-0003-2303-6519]{George H. Rieke}
\affiliation{Steward Observatory, University of Arizona,
933 North Cherry Avenue, Tucson, AZ 85719, USA}

\author[0000-0003-4528-5639]{Pablo G. P\'erez-Gonz\'alez}
\affiliation{Centro de Astrobiolog\'ia (CAB), CSIC-INTA, Ctra. de Ajalvir km 4, Torrej\'on de Ardoz, E-28850, Madrid, Spain}

\author[0000-0001-8470-7094]{Nina Bonaventura}
\affiliation{Steward Observatory, University of Arizona, 933 North Cherry Avenue, Tucson, AZ 85719, USA}

\author[0000-0003-3307-7525]{Yongda Zhu}
\affiliation{Steward Observatory, University of Arizona,
933 North Cherry Avenue, Tucson, AZ 85719, USA}

\author[0000-0003-4337-6211]{Jakob M. Helton}
\affiliation{Steward Observatory, University of Arizona, 933 North Cherry Avenue, Tucson, AZ 85719, USA}

\author[0000-0001-7673-2257]{Zhiyuan Ji}
\affiliation{Steward Observatory, University of Arizona, 933 North Cherry Avenue, Tucson, AZ 85719, USA}

\author{Jane Morrison}
\affiliation{Steward Observatory, University of Arizona, 933 North Cherry Avenue, Tucson, AZ 85719, USA}

\author[0000-0002-4271-0364]{Brant E. Robertson}
\affiliation{Department of Astronomy and Astrophysics, University of California, Santa Cruz, 1156 High Street, Santa Cruz, CA 95064, USA}

\author[0000-0002-9720-3255]{Meredith A. Stone}
\affiliation{Steward Observatory, University of Arizona,
933 North Cherry Avenue, Tucson, AZ 85719, USA}

\author{Yang Sun}
\affiliation{Steward Observatory, University of Arizona, 933 North Cherry Avenue, Tucson, AZ 85719, USA}

\author[0000-0003-2919-7495]{Christina C.\ Williams}
\affiliation{NSF's National Optical-Infrared Astronomy Research Laboratory, 950 North Cherry Avenue, Tucson, AZ 85719, USA}
\affiliation{Steward Observatory, University of Arizona, 933 North Cherry Avenue, Tucson, AZ 85719, USA}

\author[0000-0001-9262-9997]{Christopher N. A. Willmer}
\affiliation{Steward Observatory, University of Arizona, 933 North Cherry Avenue, Tucson, AZ 85719, USA}




\newcommand{\GHR}[1]{\textcolor{blue}{#1}} 
\newcommand{\IS}[1]{\textcolor{violet}{#1}} 
\newcommand{\CNAW}[1]{\textcolor{orange}{#1}} 
\newcommand{\JL}[1]{\textcolor{green}{#1}} 

\defcitealias{rieke2024}{R24}

\begin{abstract} 
The James Webb Space Telescope (JWST) is revolutionizing our view of the Universe through unprecedented sensitivity and resolution in the infrared, with some of the largest gains realized at its longest wavelengths.  We present the Systematic Mid-infrared Instrument (MIRI) Legacy Extragalactic Survey (SMILES), an eight-band MIRI survey with Near-Infrared Spectrograph (NIRSpec) spectroscopic follow-up in the GOODS-S/HUDF region.  SMILES takes full advantage of MIRI's continuous coverage from $5.6-25.5\,\mu$m over a $\sim34$~arcmin$^2$ area to greatly expand our understanding of the obscured Universe up to cosmic noon and beyond.  This work, together with a companion paper by Rieke et~al., covers the SMILES science drivers and technical design, early results with SMILES, data reduction, photometric catalog creation, and the first data release. As part of the discussion on early results, we additionally present a high-level science demonstration on how MIRI's wavelength coverage and resolution will advance our understanding of cosmic dust using the full range of polycyclic aromatic hydrocarbon (PAH) emission features from $3.3-18\,\mu$m.  Using custom background subtraction, we produce robust reductions of the MIRI imaging that maximize the depths reached with our modest exposure times ($\sim0.6 - 2.2$ ks per filter).  Included in our initial data release are (1) eight MIRI imaging  mosaics reaching depths of $0.2-18\,\mu$Jy ($5\sigma$) and (2) a $5-25.5\,\mu$m photometric catalog with over 3,000 sources.  Building upon the rich legacy of extensive photometric and spectroscopy coverage of GOODS-S/HUDF from the X-ray to the radio, SMILES greatly expands our investigative power in understanding the obscured Universe.
\end{abstract}

\keywords{infrared survey $-$ infrared astronomy $-$ galaxies: evolution $-$ Active galactic nuclei $-$ data reduction}


\section{Introduction} \label{sec:intro}

The infrared wavelength regime has proven to provide numerous powerful diagnostics for understanding galactic ecosystems over cosmic time, despite often lagging behind the optical in terms of wavelength coverage, sensitivity, and resolution \citep[e.g.][]{soifer2008}. 
The Mid-Infrared Instrument \citep[MIRI;][]{rieke2015, bouchet2015, wright2023} on the James Webb Space Telescope \citep[JWST;][]{rigby2023, gardner2023} represents a great leap forward in our ability to utilize the mid-infrared (mid-IR) in understanding the Universe.  JWST's 6.5 meter aperture affords MIRI a gain in resolution by a factor of $\sim7$ in beam diameter $-$ or $\sim50$ in beam area $-$ compared to the Spitzer Space Telescope \citep{werner2004, gehrz2007}, cutting through the confusion limit that ultimately set Spitzer's sensitivity, particularly for the Multiband Imaging Photometer \citep[MIPS;][]{rieke2004, dole2004}. MIRI is, in fact, background-limited across its full wavelength range ($5-25.8\mu$m). At $12\,\mu$m, MIRI's sensitivity is a factor of $\sim1000\times$ better than the confusion limit of the Wide-field Infrared Survey Explorer \citep[WISE;][]{wright2010}, the largest jump in JWST's capabilities over previous facilities. For an expanded discussion of these gains in the context of the history of infrared astronomy, see our companion paper, \citet{rieke2024}. 

While the paradigm shifts expected with JWST are just starting to be realized, MIRI is already far exceeding initial expectations.  Surprisingly, this has included joining JWST's near-infrared instruments\footnote{The near-infrared JWST instruments Near-Infrared Camera \citep[NIRCam;][]{rieke2023a} and the Near-Infrared Spectrograph \citep[NIRSpec;][]{jakobsen2022} are referenced in this paper.} in opening up our view of cosmic dawn. MIRI has now detected emission lines in a galaxy at $z=12.33$ \citep{zavala2024} and emission-line boosted photometry in a galaxy at $z=14.32$ (Helton et al., 2024, submitted to Nature Astronomy), the current spectroscopically-confirmed highest redshift galaxy (Carniani et al, 2024, submitted to Nature). NIRCam-dark, MIRI-only detections \citep{perez-gonzalez2024a} have opened up the possibility of a new population of high-$z$ emission line galaxies accessible to JWST.  MIRI's view of the rest-frame optical in the first $\sim500$ million years after the Big Bang \citep[see also][]{hsiao2024, calabro2024a} is providing indispensable information on the earliest galaxies. 

\begin{figure}[tbh!]
    \centering
    \includegraphics[width=1\columnwidth]{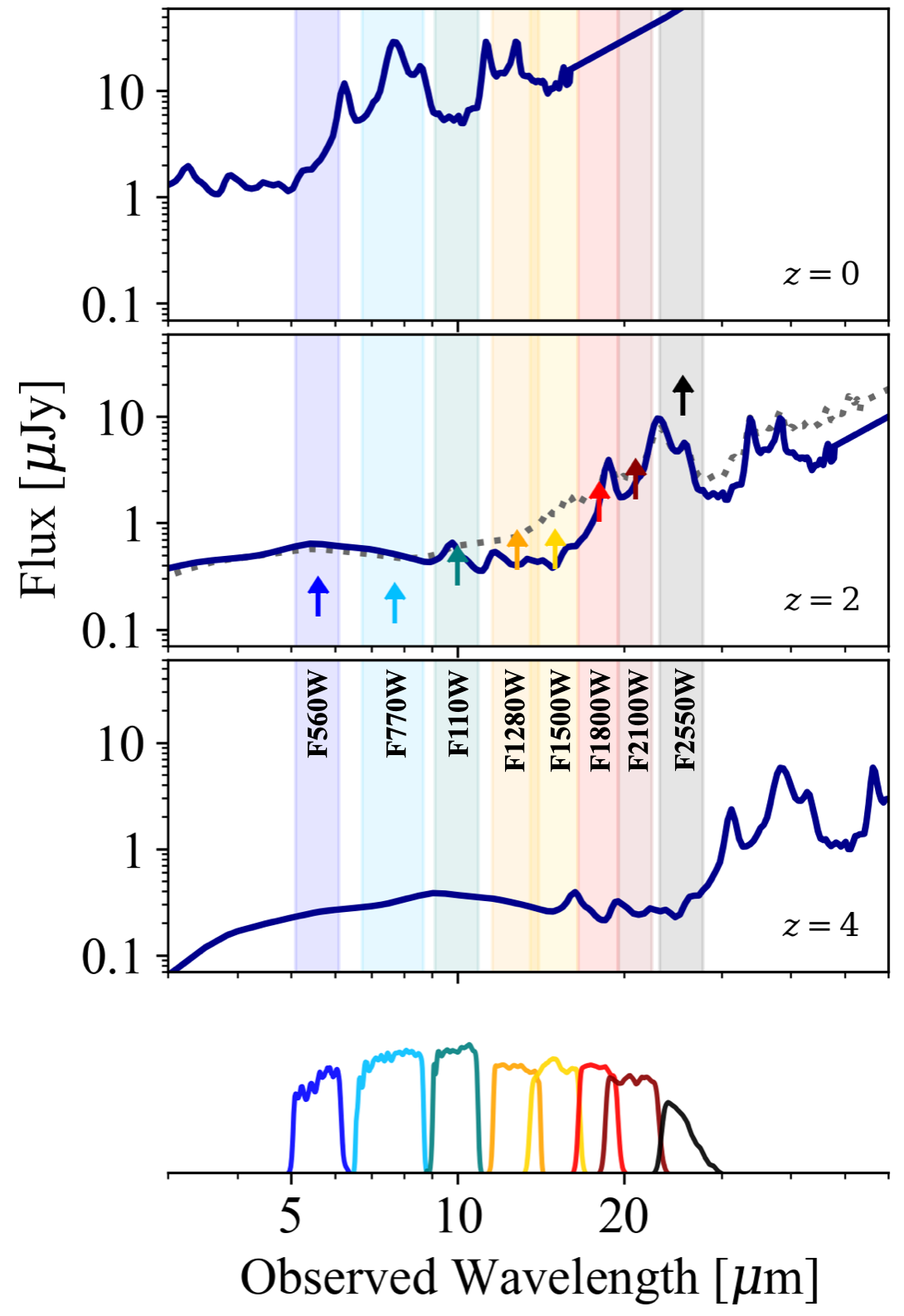}
    \caption{The SED of an infrared luminous SFG \citep{kirkpatrick2015} shown at three redshifts ($z=0,2,4$) relative to eight MIRI filters at $5.6-25.5\,\mu$m (bottom, F1130W is not shown).  From the local Universe through cosmic noon ($1<z<3$), the MIRI bands probe at least one strong mid-infrared dust (PAH) feature plus stellar continuum emission in the rest-frame near-infrared.  In the middle panel, the SFG is normalized to $10\,\mu$Jy at $21\mu$m (the original sensitivity goal for SMILES, see \citetalias{R24}) and we show the $5\sigma$ detection limits achieved by SMILES (Table~\ref{tbl:mosaic}). We additionally show a $50\%$ composite AGN spectrum \citep[gray dotted line, normalized at $21\,\mu$m;][]{kirkpatrick2015} for contrast; note the excess infrared emission at rest-frame $3-5\,\mu$m above the SFG continuum.  Beyond cosmic noon ($z\gtrsim3-4$), MIRI continues to provide an anchor on the stellar continuum and/or strong optical or near-infrared emission lines. 
    }
    \label{fig:mir}
\end{figure}

\begin{figure}[ht!]
    \centering
    \includegraphics[width=\columnwidth]{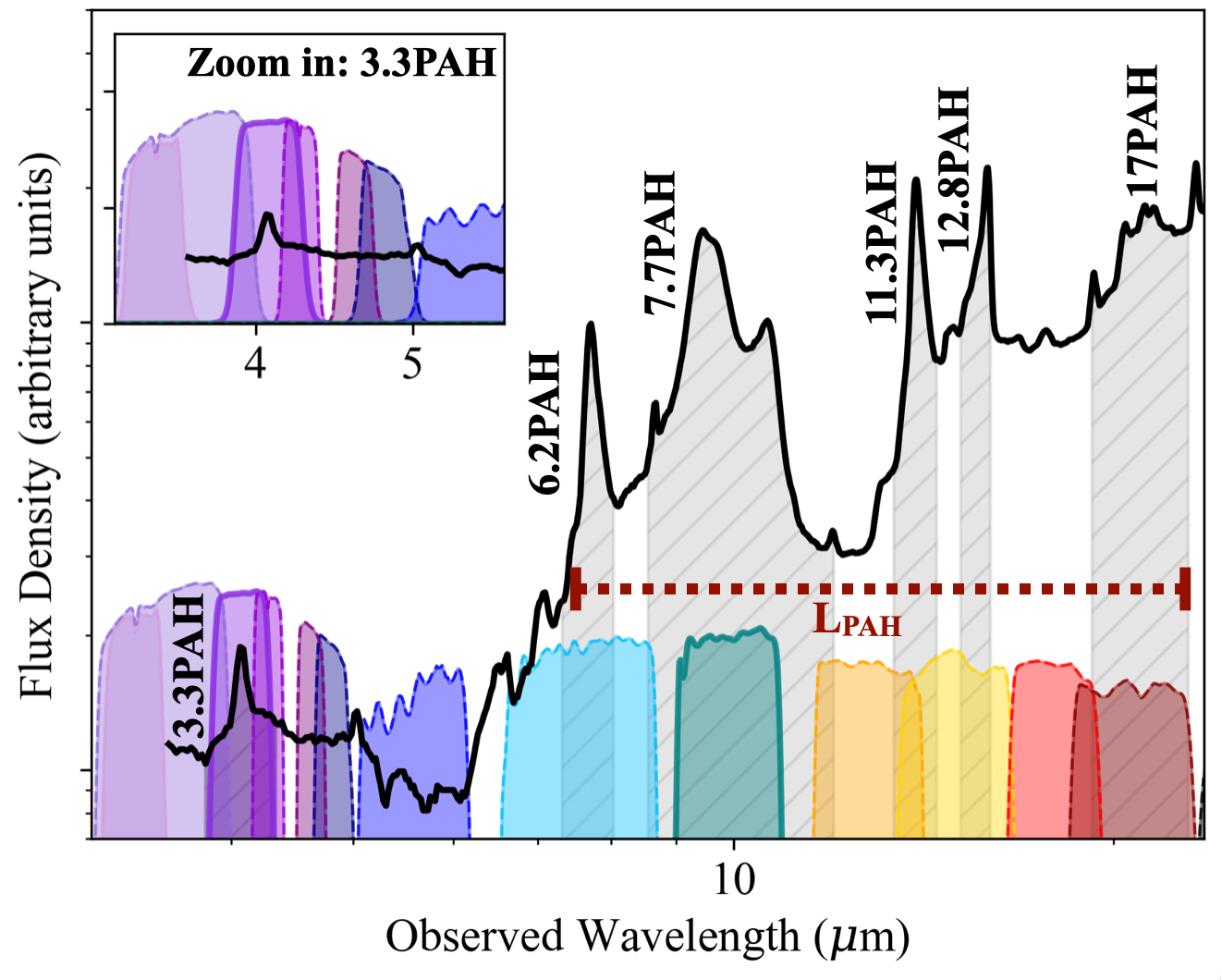}
    \caption{A typical PAH spectrum redshifted to $z=0.25$ \citep[black line;][]{lai2020}.  At low redshift, the MIRI F770W-F2550W bands cover all of the strong PAH features from $6-18\,\mu$m with $R\sim5$ resolution. The hatched regions show the PAH complex boundaries as defined in \citet{draine2021}. We define the total PAH luminosity $L_{\rm PAH}$ as the sum of these features, modeled and continuum subtracted through SED fitting. A weak dust feature, the 3.3$\mu$m PAH, is in the F410M filter (inset) and can be measured as color excess by subtracting flanking continuum bands \citep{inami2018, lai2020}.  Pre-JWST, studies of this feature were primarily limited to very massive, local galaxies \citep[e.g.][]{lai2020}. 
    }
    \label{fig:mb}
\end{figure}

At later times $-$ the originally envisioned purview of the instrument $-$ MIRI covers a large range of important spectral regimes (Figure~\ref{fig:mir}).  At $3\lesssim z\lesssim 9$, MIRI is providing rest-frame near-infrared and optical anchors (Figure~\ref{fig:mir}) in studies of dusty \citep{alvarez-marquez2023a, rodighiero2022} and/or optically dark galaxies and the so-called Little Red Dots \citep[LRDs;][]{williams2023a, wang2024a, perez-gonzalez2024}; quiescent galaxies \citep[][Ji et al, 2024, in prep]{alberts2023}; and emission line galaxies during reionization \citep[e.g.][]{alvarez-marquez2023, caputi2023, rinaldi2023, rinaldi2023a}.  During cosmic noon ($1\lesssim z\lesssim 3$) $-$ the peak epoch in star formation and black hole activity \citep{madau2014, hickox2018} $-$ the prevalence of cosmic dust means that MIRI plays a crucial role in quantifying dust-obscured activity in star-forming galaxies \citep[SFGs; e.g.][]{ronayne2023, lin2024, shivaei2024, spilker2023} and Active Galactic Nuclei \citep[AGN; e.g.][]{yang2023a, kirkpatrick2023, lyu2024}.  Furthermore, MIRI's coverage of multiple dust emission features up to this epoch (Figure~\ref{fig:mb}) will place strong constraints on the properties of small grain dust $-$ namely polycyclic aromatic hydrocarbons (PAHs) $-$ which are known to be correlated with the (obscured) star formation rate \citep[e.g.][]{rujopakarn2013, shipley2016, shivaei2024} and likely molecular gas \citep[e.g.][]{cortzen2019}.  
    
In this paper, we present the Systematic Mid-infrared Instrument Legacy Extragalactic Survey (SMILES), the largest and thus far only survey that includes MIRI imaging in eight bands,
providing full wavelength coverage over the $5.6-25.5\,\mu$m range. This multi-band coverage allows SMILES data to be applied to many scientific studies not equally accessible to surveys covering only 2 or 3 bands at wavelengths longer than $8\,\mu$m. SMILES is a blank field MIRI imaging survey with follow-up NIRSpec spectroscopy optimized for studying galaxy populations at cosmic noon and beyond in the GOODS-S/HUDF field \citep{giavalisco2004, beckwith2006}. 

The historical context and scientific motivations for SMILES and its legacy among MIRI surveys are presented in a companion paper \citet{rieke2024}; in this work, we present an overview of the ongoing science with SMILES, the technical design, data reduction, photometric catalog, and first data release of SMILES, which includes the MIRI imaging (NIRSpec spectra will be made public in a future release). In Section~\ref{sec:science}, we provide an overview of the scope and early results of SMILES, as well as a small science demonstration on the power of MIRI in probing PAH features. Sections~\ref{sec:design}, \ref{sec:reduction} and \ref{sec:catalog} describe the data acquisition, reduction, and photometric extraction of the MIRI imaging, respectively. In Section~\ref{sec:nirspec}, we overview the SMILES NIRSpec follow-up observations to the MIRI imaging. Section~\ref{sec:catalog_and_release} concludes by discussing the properties of the SMILES catalog and describes the data release.  All magnitudes in this work are in AB units \citep{oke1983}.

\begin{figure*}[tbh!]
    \centering
    \includegraphics[width=1.6\columnwidth]{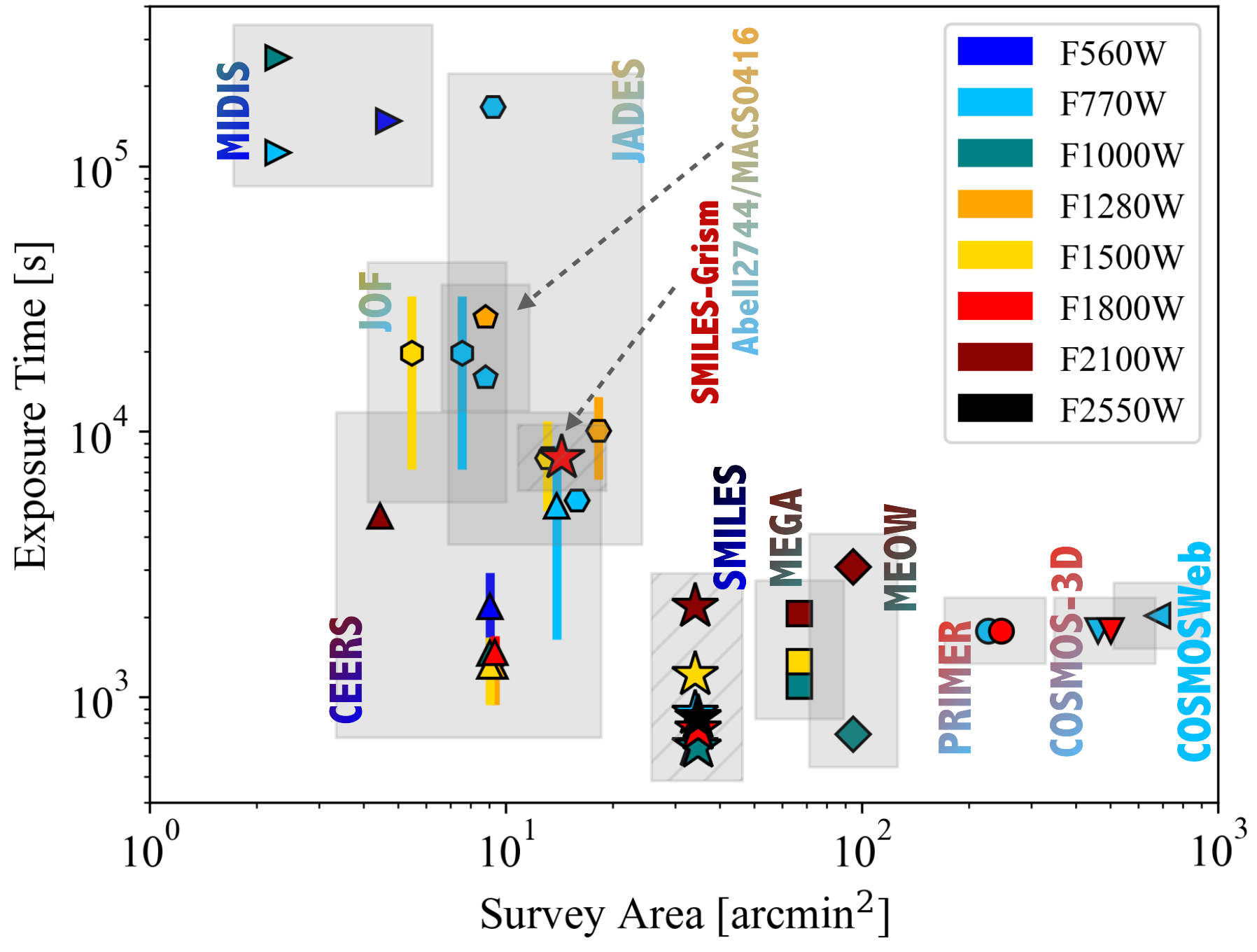}
    \caption{The distribution of exposure times as a function of survey area and filter for current and future MIRI surveys (with MIRI as prime or in parallel), through Cycle 3 (roughly from deep and narrow to shallow and wide): MIRI Deep Imaging Survey (MIDIS; PIDs 1283, 6511, PI: G. Ostlin) [right triangles]; JADES MIRI parallels \citep[Alberts et al., 2024, in prep,][]{eisenstein2023} and JADES Origin Field \citep[JOF;][]{eisenstein2023a} [hexagons]; Abell2744/MACS0416 (PID 5578, PI E. Iani) [pentagons]; CEERS \citep[][]{yang2023} [upright triangles], SMILES (this work) and SMILES-Grism (PID 4549, PI G. Rieke) [stars]; MIRI EGS Galaxy and AGN Survey (MEGA; PID 3794, PI A. Kirkpatrick) [squares]; the MIRI Early Obscured-AGN Wide Survey (MEOW; PID 5407, PI G. Leung) [diamonds]; PRIMER (PID 1837, PI J. Dunlop) [circles]; COSMOS-3D (PID 5893, PI K. Kakiichi) [downward triangles], and COSMOSWeb \citep[][]{casey2022} [left triangle].  SMILES is currently the only survey to take full advantage of the MIRI wavelength coverage.}
    \label{fig:miri_surveys}
\end{figure*}

\section{Science with SMILES}\label{sec:science}

A full treatment of the original science cases that motivated the SMILES design is presented in our companion paper, \citet{rieke2024} (hereafter R24). In this section, we provide a brief summary of the scope of the survey, early results, and a small science demonstration.

\subsection{Scope and Design}\label{sec:scope}

As demonstrated in Figure~\ref{fig:mir}, the MIRI imaging bands cover multiple prominent spectral features, spanning the rest-frame near- to mid-infrared up to cosmic noon ($1<z<3$). Up to this epoch, the primary type of infrared source is massive star-forming galaxies, whose SEDs are dominated by broad PAH features in MIRI's wavelength range. At the peak in activity at cosmic noon, Luminous Infrared Galaxies (LIRGs, $L_{\rm IR}=10^{11-12}\,\Lsun$) forming stars in dust-rich environments account for the bulk of the star formation density \citep[e.g.][]{lefloch2005, perez-gonzalez2005} and commonly reside on the main sequence of SFGs \citep[e.g.]{elbaz2011, magnelli2011, noeske2007}, in sharp contrast to their rarity in the local Universe \citep[e.g.][]{sanders2003}.  In addition, obscured AGN are a prominent infrared population at cosmic noon, whose intrinsic near- to mid-IR SEDs are dominated by emission from warm dust that manifests as infrared continuum excess above the stellar continuum and/or PAH features (Figure~\ref{fig:mir}, middle panel). It's been well established that the bulk of black hole growth during this epoch is obscured \citep[e.g.][]{delmoro2016, padovani2016, ananna2019}; however, pre-JWST, our ability to identify obscured AGN was limited \citepalias{rieke2024}.  MIRI is uniquely suited among the JWST instruments to provide unbiased samples of massive star-forming galaxies and AGN in this critical era.

\begin{figure*}[ht!]
    \centering
    \includegraphics[width=1.8\columnwidth]{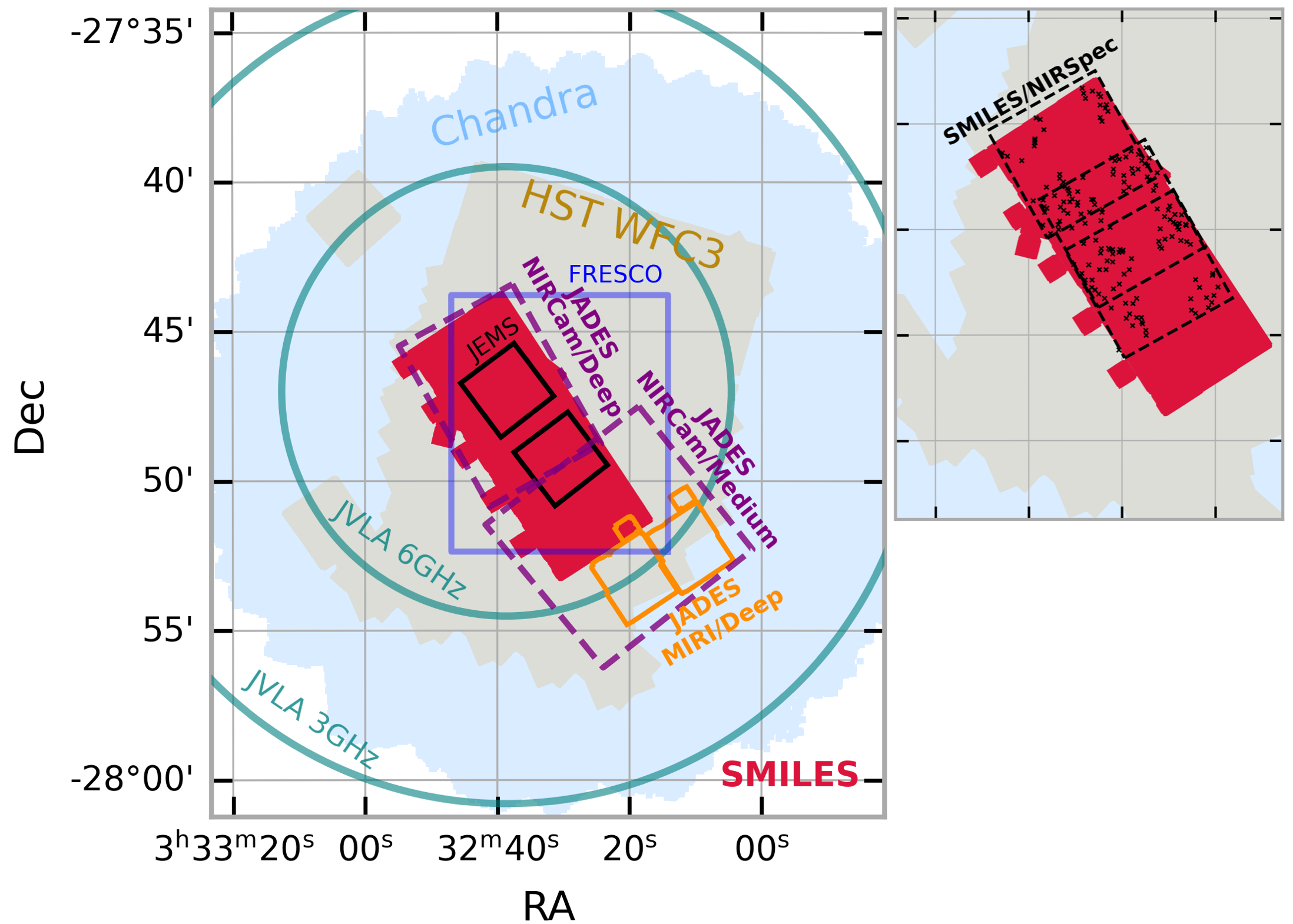}

    \caption{(left) The SMILES MIRI imaging footprint (red) in the GOODS-S/HUDF region relative to a subset of the ancillary datasets available: JADES/NIRCam Deep and Medium imaging \citep{rieke2023, eisenstein2023} and JADES/MIRI Deep parallels (Alberts et al., 2024, in prep), JEMS NIRCam medium band imaging \citep{williams2023}, FRESCO NIRCam grism and imaging \citep{oesch2023}, the HST WFC3 F160W footprint \citep{grogin2011}, JVLA 6 and 3 GHz imaging out to the half-power radius \citep{alberts2020}, and Chandra 7 Ms X-ray imaging \citep{luo2017}. (right) A zoom-in of the SMILES field showing the three SMILES/NIRSpec MSA pointings with targets assigned to slits marked (see Section~\ref{sec:nirspec}).
    }

    \label{fig:footprint}
\end{figure*}

\begin{figure*}[tbh!]
    \centering
    \includegraphics[width=1.05\textwidth]{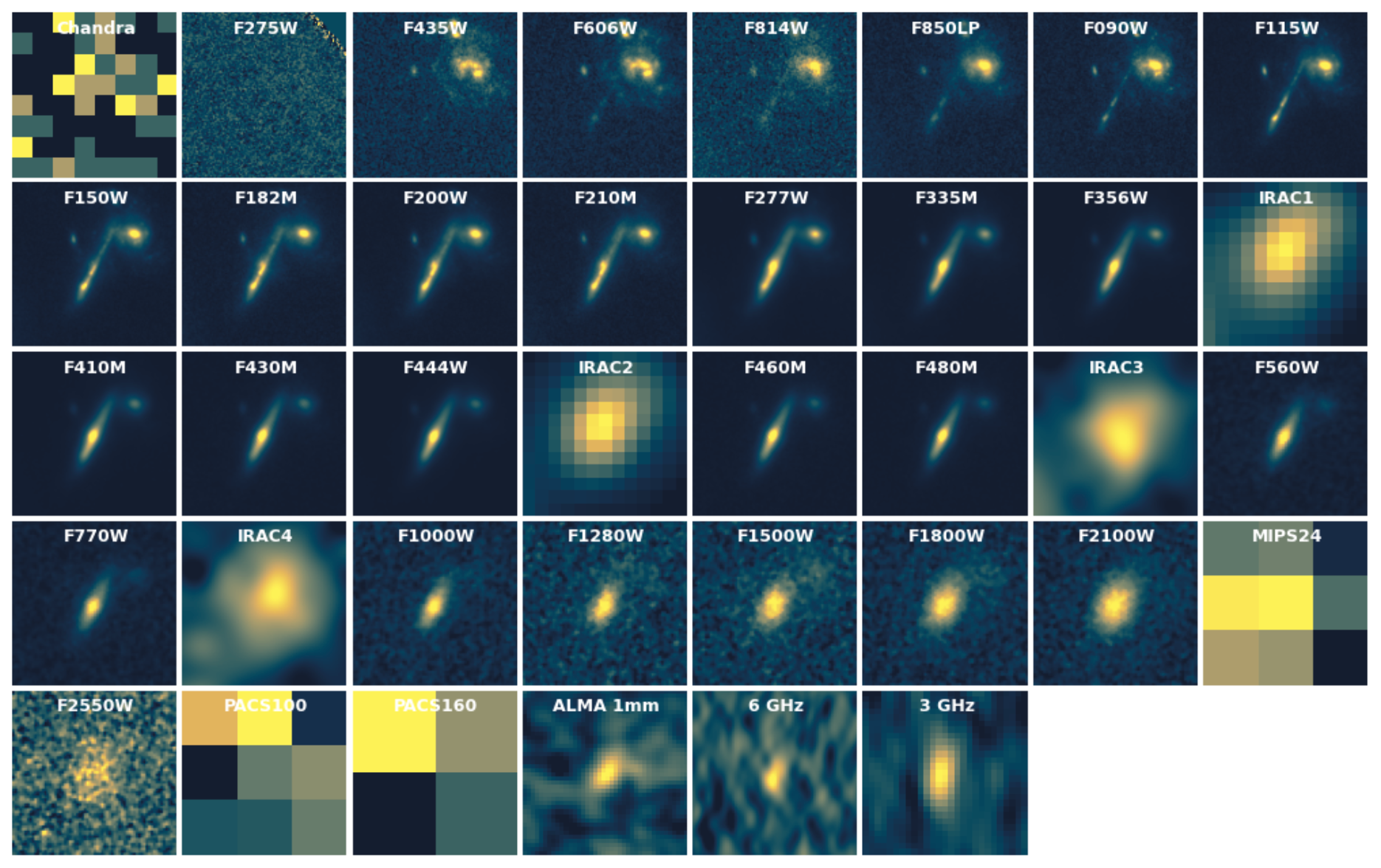}
    \caption{A $z=1.76$ galaxy shown in 38 bands of photometry, from X-ray to radio, in $4\arcsec\times4\arcsec$ cutouts. The dramatic increase in resolution and sensitivity provided by MIRI  relative to Spitzer/IRAC and MIPS is demonstrated in this figure; bulge and disk features are clearly seen in the MIRI bands.  A line-of-sight companion grand design spiral at $z\sim1.3$ to the upper right is apparent in HST, NIRcam, and MIRI F560W/F770W, then fades into the longer MIRI bands, indicating low dust content. We additionally include Herschel/PACS 100 and $160\mu$m to demonstrate that this gain has no equivalent in the far-infrared, leaving a large gap between JWST and ALMA.}
    \label{fig:photometry}
\end{figure*}

SMILES was thus designed to take a census of and investigate the nature of these luminous infrared populations at $1<z<3$ through MIRI imaging (see \citetalias{rieke2024} for more details) with follow-up NIRSpec Multi-Object Spectroscopy \citep[MOS;][]{ferruit2022, rawle2022}.   
SMILES covers $34.5$ arcmin$^2$ centered on the GOODS-S/HUDF field with MIRI imaging in eight bands, with integration times and achieved sensitivities listed in Table~\ref{tbl:mosaic}. As shown in Figure~\ref{fig:miri_surveys}, SMILES occupies a parameter space similar to other wide, modest depth MIRI surveys but with the distinction of continuous wavelength coverage from $5.6-25.5\,\mu$m. 
To date, the only other blank-field MIRI survey, current or in Cycle 3, with imaging in more than three bands is CEERS \citep{yang2023}.  Equally important  \citepalias{rieke2024} is the ancillary data available; the SMILES footprint is shown in Figure~\ref{fig:footprint} compared to highly complementary datasets.  GOODS-S/HUDF was chosen to maximize the quality and quantity of ancillary data needed for the survey goals. This includes ultra-deep NIRCam imaging from the JWST Advanced Deep Extragalactic Survey \citep[JADES;][]{rieke2023, eisenstein2023} and the JWST Extragalactic Medium-band Survey \citep[JEMS;][]{williams2023}, HST imaging from multiple surveys \citep[e.g.][]{grogin2011}, deep ALMA imaging and spectroscopy \citep{walter2016, dunlop2017, hatsukade2018}, ultra-deep imaging surveys in the X-ray with Chandra \citep{luo2017}  and radio with VLA \citep{alberts2020}, and copious spectroscopy \citep{momcheva2016, bacon2023, bunker2023}, including NIRCam grism spectroscopy from the First Reionization Epoch Spectroscopically Complete Observations survey \citep[FRESCO;][]{oesch2023}.  Including  JWST coverage, $\gtrsim30$ bands of high-resolution photometry from the UV$-$radio are available over the SMILES footprint, as demonstrated in Figure~\ref{fig:photometry}.  In the next section, we give an overview of some of the early science results from SMILES, along with some of the future directions they suggest.

\begin{table*}[ht]
    \centering
    \begin{tabular}{lccccccc}
    \hline
    \hline
        Filter & N$_{\rm groups}$ & N$_{\rm int}$ & Exp Time & Aperture Radius & $5\sigma$ Limit & ETC v3.0 & Improvement  \\
         &  &  & s & (65\% EE) & $\mu$Jy (AB) & $\mu$Jy & Factor \\
         (1) & (2) & (3) & (4) & (5) & (6) & (7) & (8) \\
    \hline

    F560W & 59 & 1 & 654.9 & 0.48$^{\prime\prime}$ & 0.21 (25.6) & 0.46 &  2.2 \\
    F770W & 78 & 1 & 865.8 & 0.42$^{\prime\prime}$ & 0.20 (25.7) & 0.52 & 2.6 \\
    F1000W & 58 & 1 & 643.8 & 0.36$^{\prime\prime}$ & 0.39 (24.9) & 0.83 & 2.1 \\
    F1280W & 68 & 1 & 754.8 & 0.42$^{\prime\prime}$ & 0.62 (24.4) & 1.3 & 2.1 \\
    F1500W & 101 & 1 & 1121.1 & 0.49$^{\prime\prime}$ & 0.75 (24.2) & 1.4 & 1.9 \\
    F1800W & 68 & 1 & 754.8 & 0.54$^{\prime\prime}$ & 1.8 (23.3) & 3.1 & 1.7 \\
    F2100W & 32 & 6 & 2186.7 & 0.60$^{\prime\prime}$ & 2.8 (22.8) & 3.1 & $-$\\
    F2550W & 18 & 4 & 832.5 & 0.72$^{\prime\prime}$ & 17 (20.8) & 18 & $-$ \\
    \hline
    \end{tabular}
    \caption{The exposure setup and depths of the SMILES mosaics. Notes: Column 4: Total exposure time over four dithers. Each pointing was observed for a total of 2.17 hours of science time including all filters. Columns 5-8: The $5\sigma$ point source sensitivity  measured from our data (column 6) compared to ETC v3.0 predictions (column 7).  The point source sensitivity was measured using randomly placed circular apertures with a radius equal to the $65\%$ encircled energy (EE; column 5) of our PSFs (Section~\ref{sec:photometry}) with aperture corrections applied.  The ETC predictions were obtained using the same size aperture and a $1-2\arcsec$ ($1.5-2.5\arcsec$ at $\geq21\mu$m) annulus to estimate background sky and noise, with the background corresponding to Dec.$\,15$, 2022 in GOODSS.  Column 8: The ratio of our measured $5\sigma$ point source sensitivity to the ETC prediction.}
    \label{tbl:mosaic}
\end{table*}

\begin{figure*}[ht!]
    \centering
    \includegraphics[width=1.9\columnwidth]{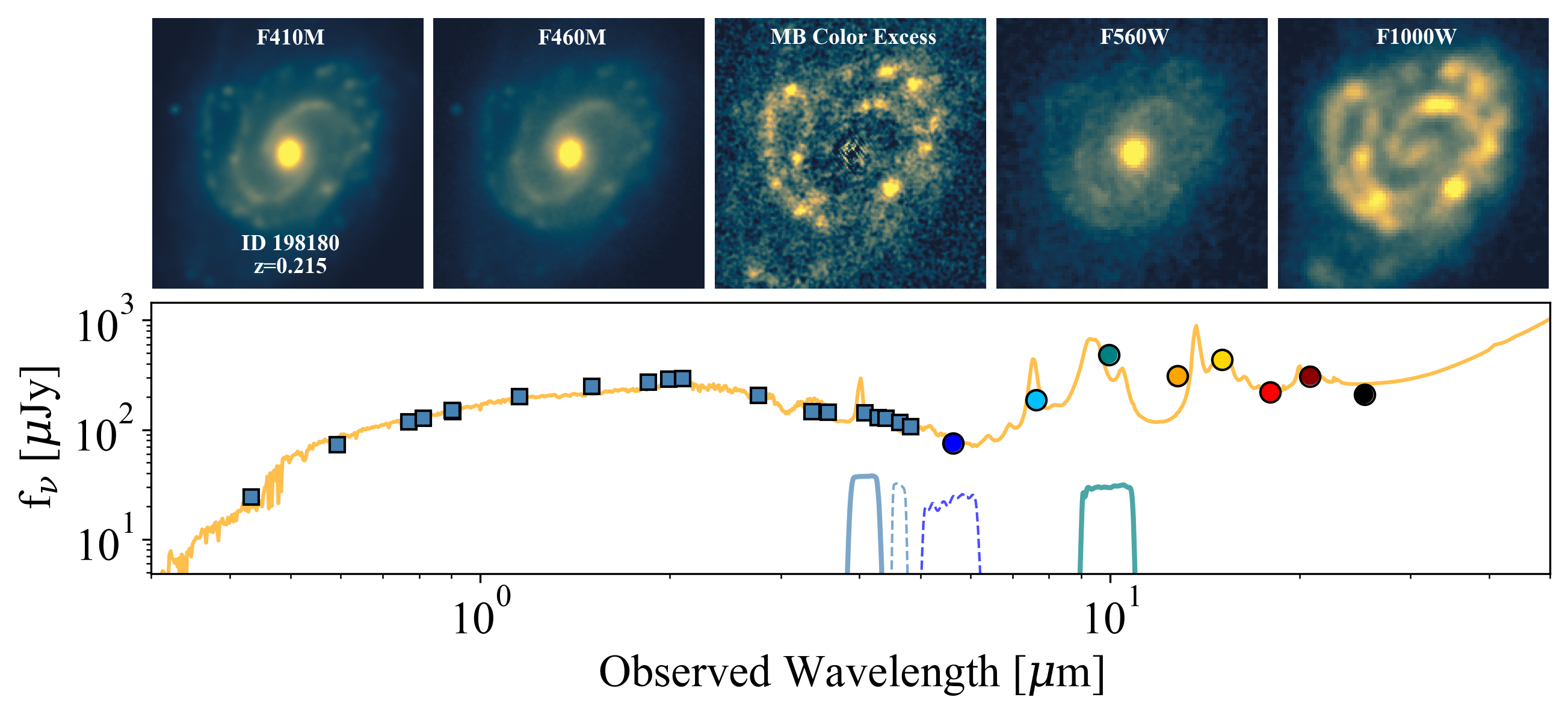}
    \caption{Cutouts (top) and SED (bottom) of a $z\sim0.2$ star-forming galaxy where the MIRI bands provide excellent sampling of the strong PAH features and the weaker, poorly studied 3.3PAH is in the NIRCam medium band F410M (solid blue filter).  The medium-band color excess image (middle cutout) shows that the 3.3PAH lights up in the same spatial regions as the 7.7PAH complex, which is wide enough to dominate the flux observed in F1000W (solid teal filter).  Filters containing emission features are shown as solid lines, continuum filters with dashed lines.  Errorbars are smaller than the datapoints.
    }
    \label{fig:pah_cutouts}
\end{figure*}

\subsection{Early Results with SMILES}\label{sec:miri_mot}

\subsubsection{Obscured AGN}\label{sec:agn_science}

A new census of AGN in GOODSS-S has recently been presented in \citet{lyu2024}, identifying 111 AGN in massive ($M_* > 10^{9.5} M_\odot$) galaxies with the inclusion of SMILES data, $34\%$ of which were previously unidentified.  This increase is remarkable if one considers the uniquely extensive datasets in the GOODS-S/HUDF region.  \citet{lyu2024} evaluated the full range of AGN search techniques, finding largely independent contributions from infrared SED fitting (e.g. including SMILES); selection via X-ray luminosity and X-ray to radio ratios; variability; radio loudness and flat spectrum radio emission; and optical spectroscopy. Through this, they confirmed previous arguments that a panchromatic approach is necessary to take a census of AGN with high completeness \citep[e.g.][]{donley2008, delvecchio2017, alberts2020, lyu2022a}.

Beyond the above, there are a number of new questions raised by \citet{lyu2024}.  First, a comparatively large sample of AGN candidates was found in low-mass galaxies \citep[see also][]{yang2023a}.  This is reminiscent of the discovery of low-luminosity AGN in nearby dwarf galaxies \citep{reines2013, reines2020, moran2014} and may have important implications for our understanding of the role of AGN in galaxy evolution.  However, confirming the nature of these sources will be challenging due to their faint nature and uncertainties in the intrinsic SEDs of both low-luminosity AGN and their low-mass, low-metallicity hosts.  A second discovery is the identification of a few dozen AGN candidates at $z>4$ with SMILES, likely mostly obscured and in many cases very weak in the X-ray.  Again confirmation and expansion of this population in conjunction with other ongoing efforts to study AGN at high-$z$ \citep[e.g.][]{maiolino2023, harikane2023, ubler2023} will be important to constrain AGN evolution.

\subsubsection{PAH features to cosmic noon and beyond}\label{sec:pah_science}

The multiple PAH features prominent in the mid-infrared (Figures~\ref{fig:mir}$-$\ref{fig:mb}) provide a wealth of information on small-grain dust properties, are a proxy for obscured star formation \citep[e.g.][]{shipley2016}, and constrain the broader role of dust in the chemistry and energy exchange in the Interstellar Medium \citep[ISM; e.g.][]{mckinney2021}.  An initial study of PAH features in $0.7<z<2$ SFGS has been presented by \citet{shivaei2024}, which demonstrates that the MIRI bands are powerful tracers of these features, probing at least one major PAH feature up to $z\sim2.5$ if F2100W imaging is available.  In that work, the PAH mass fraction (q$_{\rm PAH}$), obscured SFR, and obscured fraction are quantified in 443 galaxies down to SFRs of $<1\,\Msun$ yr$^{-1}$ and masses $\sim10^9\,\Msun$.  This SFR limit is over an order of magnitude lower than what was achievable at cosmic noon with Spitzer \citep{elbaz2011,magnelli2011,reddy2008,shipley2016,shivaei2015,shivaei2017}, showing that even at moderate depth, MIRI can reach galaxies on and below the main sequence during cosmic noon. 

This work highlights open questions we can now address. (1) Looking at the trend of PAH mass fraction with decreasing metallicity via the mass-metallicity relation \citep{sanders2021}, \citet{shivaei2024} found that the local q$_{\rm PAH}$-metallicity trend appears to hold at cosmic noon. However, the modest sample of robust detections at very low metallicity in SMILES motivates deeper multi-band MIRI surveys combined with spectroscopy to directly measure metallicities, a goal of our NIRSpec follow-up (Section~\ref{sec:nirspec}).  (2) The mid-infrared can now be calibrated as a SFR indicator down to sub-LIRG main sequence galaxies at cosmic noon (Alberts et al., 2024b, in prep). And (3), JWST opens up avenues for understanding the diagnostic power of tracing multiple PAH features simultaneously; we revisit this briefly in Section~\ref{sec:demo}.

\subsubsection{Morphology of obscured star formation}\label{sec:pah_science2}

The ability of MIRI to spatially resolve mid-infrared emission up to cosmic noon deserves special mention.  Figure~\ref{fig:photometry} demonstrates the gain in resolution in MIRI, a $\sim7\times$ improvement over previous facilities and only a factor of $\sim2$ coarser resolution at 7.7$\,\mu$m than NIRCam at $4.4\,\mu$m.  Figure~\ref{fig:pah_cutouts} shows a particularly spectacular example: spatially resolved PAH emission in F1000W in a galaxy at $z\sim0.3$ (see Section~\ref{sec:demo}).  Taking advantage of SMILES multi-band imaging, a spatially resolved study of the 6.2$\,\mu$m PAH is now possible up to $z\sim1.5$. The critical advantage SMILES here is that multiple bands allow the selection of the filter containing the PAH band at different redshifts, maintaining the highest possible diffraction-limited angular resolution (Florian et al., 2024, in prep).  This work is finding that high redshift LIRGs have extended star formation, as traced by the PAH via MIRI, and less disturbed morphologies (as traced by the stellar mass distribution via NIRCam) than LIRGs in the local Universe.  This points to a disparate evolution at cosmic noon, where luminous dusty starbursts are not triggered by major mergers as is the case locally and as predicted in a common framework for massive galaxy evolution \citep[e.g.][]{hopkins2008}.  Instead minor interactions and gas accretion from the Intergalactic Medium (IGM) likely play bigger roles.

\subsubsection{Optical dark galaxies and little red dots}\label{sec:lrd_science}

Pushing past the original design scope of SMILES \citetalias{rieke2024}, \citet{williams2023a} and \citet{perez-gonzalez2024} used SMILES data as part of studies of optically dark galaxies and Little Red Dots \citep[LRDs; e.g.][]{matthee2023, labbe2023a, furtak2023} at $3\lesssim z\lesssim7$.  At these redshifts, MIRI provides the rest-frame near-infrared (Figure~\ref{fig:mir}), which is needed to robustly measure the properties of very red galaxies \citep{williams2023a}.  For LRDs in particular, NIRCam imaging presented strong evidence for reddened AGN \citep[e.g.][]{labbe2023a} and broad-line AGN have been confirmed in some LRDs \citep[e.g.][]{greene2023, wang2024}.  The behavior of the LRD SEDs at observed wavelength of 10 $\mu$m and longer with SMILES, however, very clearly shows a rollover, indicative of substantial contribution by stellar photospheric emission, both on average via stacking  \citep{williams2023a} and in individual galaxies \citep{perez-gonzalez2024}.  The latter is required for exploring this unknown and possibly heterogeneous population; however, SMILES is only deep enough to obtain good SED constraints for a few examples. A follow-up program to obtain much deeper MIRI observations is needed.  

\subsection{PAHs at low redshift: a small science demonstration}\label{sec:demo}

As discussed in Sections~\ref{sec:pah_science}-\ref{sec:pah_science2}, the MIRI bands are a powerful tracer of mid-infrared dust emission, probing at least one strong PAH feature up to $z\sim2.5$.  At low redshift ($z\lesssim0.3-0.6$), MIRI photometry covers all strong PAH features at 6.2, 7.7, 11.3$\,\mu$m and even the generally weaker features at 12.8 and 17$\,\mu$m (Figure~\ref{fig:mb}) with a spectral sampling equivalent to $R = \lambda/\Delta\lambda \sim 5$.  This capability is a powerful complement to MIRI spectroscopy, which will provide an unmatched look at small grain dust physics and the role of dust within the ISM \citep[e.g.][]{garcia-bernete2022a, young2023} for small samples, given the small FOV and/or limited wavelength coverage of the MIRI Medium Resolution Spectrometer \citep[MRS;][]{wells2015, argyriou2023} and Low Resolution Spectrometer \citep[$5-14\,\mu$m, LRS;][]{kendrew2015}.

Here we provide a high-level demonstration of the parameter space opened by MIRI+NIRCam using 15  $z_{\rm spec}\sim0.2-0.5$ (log $\logM=8-11$) star-forming galaxies\footnote{AGN have been removed from the sample using the catalog in \citet{lyu2024}.} where MIRI covers the major PAH features at $5-18\,\mu$m.  
We fit their HST, NIRCam, and MIRI photometry using {\tt Bagpipes} using a non-parametric star formation history, flexible attenuation curve, and dust emission parameterized based on the \citep{draine2021} models.  An example SED and cutouts are shown in Figure~\ref{fig:pah_cutouts} for a galaxy at $z\sim0.2$ where the F1000W filter is dominated by the 7.7PAH and has a distinctive morphology compared to F560W, which is showing stellar continuum.  The well-sampled mid-IR SED allows us then to estimate the continuum-subtracted luminosity in these PAH features: $L_{\rm PAH}\equiv L_{\rm 6.2+7.7+11.3+12.8+17}$.  

At this redshift, a weaker, but significant PAH feature, the $3.3\,\mu$m PAH, can be extracted from the NIRCam medium bands from JADES and JEMS as a color excess relative to continuum-dominated filters (Figure~\ref{fig:mb}, inset). Due to being a narrow, weak feature, pre-JWST the 3.3PAH was only accessible in luminous galaxies in the local Universe and is still poorly understood \citep{sandstrom2023}.  As a feature composed primarily of very small, neutral grains \citep[e.g.][]{maragkoudakis2020}, the 3.3PAH may provide useful constraints on dust physics when combined with the stronger features and may be preferentially destroyed in harsh radiation fields, a controversial topic being explored by multiple local JWST surveys \citep[e.g.][]{schroetter2024}. In addition, the 3.3PAH has particular importance to high redshifts studies with JWST, as it is the only dust emission feature accessible at $z\gtrsim3$ \citep{spilker2023}. 

\begin{figure}
    \centering
    \includegraphics[width=\columnwidth]{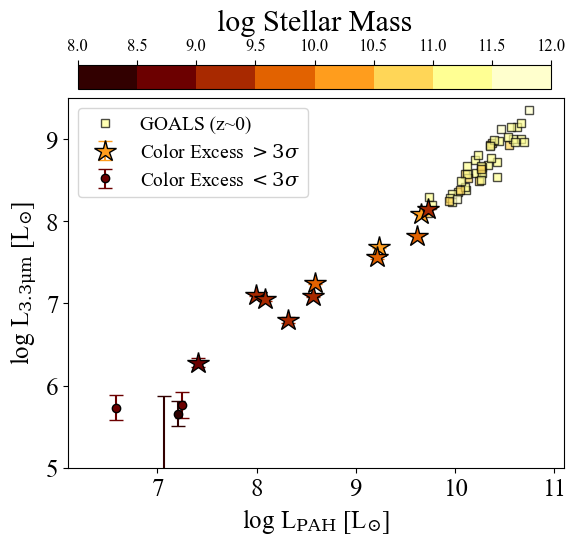}
    \caption{The luminosity of the 3.3PAH measured from NIRCam medium-bands as a function of the total PAH luminosity as estimated from SED fitting of MIRI photometry, which covers 5 major PAH features at $z\sim0.2-0.5$ (Figure~\ref{fig:pah_cutouts}).  Stars (circles) show color excess in the medium bands measured at $>3\sigma$ ($<3\sigma$).  The squares show the GOALS $z\sim0$ LIRG sample.  The color bar shows the stellar mass; our demonstration confirms that the 3.3PAH scales with stronger PAH features in star-forming galaxies at higher redshift and lower mass than previously possible. }
    \label{fig:pah33}
\end{figure}

In Figure~\ref{fig:pah33}, we show the luminosity of the 3.3$\,\mu$m PAH\footnote{We measure the 3.3PAH luminosity by interpolating the continuum under the line based on flanking filters close to medium bands that contain the emission feature based on spectroscopic redshifts.  The relevant medium band is F410M, F460M, or F480M for our sample.  Subtracting the continuum, we measure a color excess, from which we derive a line flux based on the medium band filter bandwidth and the throughput at the observed wavelength of the line.} relative to $L_{\rm PAH}$ and the stellar mass derived from SED fitting.  Our sample extends the linear correlation seen at $z\sim0$ in the GOALS sample \citep{lai2020} to both higher redshifts and lower masses than previously possible.  Furthermore, in Figure~\ref{fig:pah_cutouts}, difference images showing the color excess demonstrate that the spatial distribution of the 3.3PAH is qualitatively similar to the 7.7PAH in the MIRI F1000W filter. This high-level demonstration of the SMILES data shows the parameter space opened by MIRI imaging.  A full spatially-resolved analysis of this sample will be presented in future work. A spectroscopic analysis of the 3.3PAH using NIRCam grism from the FRESCO survey will be presented in Lyu et al., 2024, in prep.

\section{Technical Design and Data Acquisition}\label{sec:design}

SMILES (PID 1207; PI: G. Rieke) consists of 8 mosaics (filters F560W, F770W, F1000W, F1280W, F1500W, F1800W, F2100W, and F2550W) built from 15 MIRI pointings in a $5\times3$ configuration centered on $\alpha,\delta$=3:32:34.8,-27:48:32.6 (J2000, Figure~\ref{fig:footprint}).  The center and the aperture position angle of 33 degrees were chosen to maximize overlap with the projected JADES NIRcam imaging (which was successfully achieved) and enable scheduling during a period of low background.  The majority of the MIRI imaging was obtained in Dec 2022, save for two tiles (2 and 6; see Figure~\ref{fig:exp_time}) that suffered from guide star failures shortly into observations.  These pointings were reobserved in all filters in Jan 2023; see Appendix~\ref{app:a} for further details.  The final area covered is $\sim34.5$ arcmin$^2$ with $5\%$ overlap between tiles, ensuring relatively uniform minimum exposure time across the mosaics (Figure~\ref{fig:exp_time}). As discussed in the previous section, the exposure setup was aimed at the analysis of AGN and SFGs at cosmic noon, which we projected would require exposure times ranging from $\sim650-2100$ s corresponding to $5\sigma$ limits of 0.25-4 $\mu$Jy via ETC v1.0 for F560W - F2100W (Table~\ref{tbl:mosaic}).  For F2550W, where the thermal background of the telescope is very high, we aimed at detecting galaxies at $z\sim1$ down to $25\,\mu$Jy.  This resulted in a total of 2.17 hr of science time per pointing.  Due to the relatively short exposure times in each band, we opted not to add coordinated parallel observations.

To maximize sensitivity via long ramps, we opted for single integrations except for F2100W and F2550W where the background would saturate.  Standard 4-pt dithers (optimized for point sources to maximize area with uniform exposure time) were chosen to ensure adequate PSF sampling and minimize detector artifacts and cosmic ray hits/showers (Section~\ref{sec:stage1}) and to improve background subtraction (Section~\ref{sec:background}). The recommended readout mode FASTR1 was used and filters were observed in order of increasing wavelength to mitigate persistence \citep{dicken2024}. Total exposure times and the breakdown into groups and integrations are shown in Table~\ref{tbl:mosaic}.

\section{Data Processing}\label{sec:reduction}

This section describes the data processing for SMILES, which is broadly similar to the reductions described in previous works \citep[e.g.][]{yang2023, morrison2023, perez-gonzalez2024}.  Data reduction is done using v1.12.5 of the {\tt JWST Calibration Pipeline} \citep{bushouse2023} with CRDS\footnote{Calibration Reference Data System: \url{https://jwst-crds.stsci.edu/}} version 11.17.7 and CRDS context 1188.  Custom external routines are used for identifying warm pixels, background subtraction, and astrometry correction at appropriate stages of the pipeline. We start with the uncalibrated ({\tt \_uncal}) files obtained from MAST\footnote{\url{https://mast.stsci.edu/portal/Mashup/Clients/Mast/Portal.html}}.

\subsection{Stage 1: Detector Processing}\label{sec:stage1}

Stage 1 ({\tt calwebb\_detector1}) of the JWST pipeline performs detector-level processing on raw ramp data, including corrections for non-linearity, reset anomaly, first and last frame effects, dark subtraction, and jump detection specific to the MIRI imager.  Based on pre-flight ground-based and commissioning data \citep[see also][for MIRI data reduction simulations]{yang2021}, we adopt the default parameters in the pipeline for this step; a detailed discussion of the non-ideal behaviors and the corrections implemented in stage 1 of the pipeline is presented in \citet{morrison2023}.  As noted in that work, several artifacts are not currently addressed in the pipeline, including persistence, the cruciform feature of the PSF at F560W and F770W, interpixel capacitance, column and row striping, the first exposure effect, and the brighter-fatter effect \citep{argyriou2023a}.  As noted in Section~\ref{sec:design}, we observed each of our pointings in order of increasing wavelength to mitigate persistence. We address the cruciform and row/column stripping in Sections~\ref{sec:background} and \ref{sec:photometry}, respectively, and we discuss here the recent implementation in the pipeline of a correction for large cosmic ray events, called `cosmic ray showers' due to their ability to affect a large number of pixels on the MIRI detectors. We expect the other effects to be minimal in our data. 

\begin{figure}[h!]
    \centering
    \includegraphics[width=0.95\columnwidth]{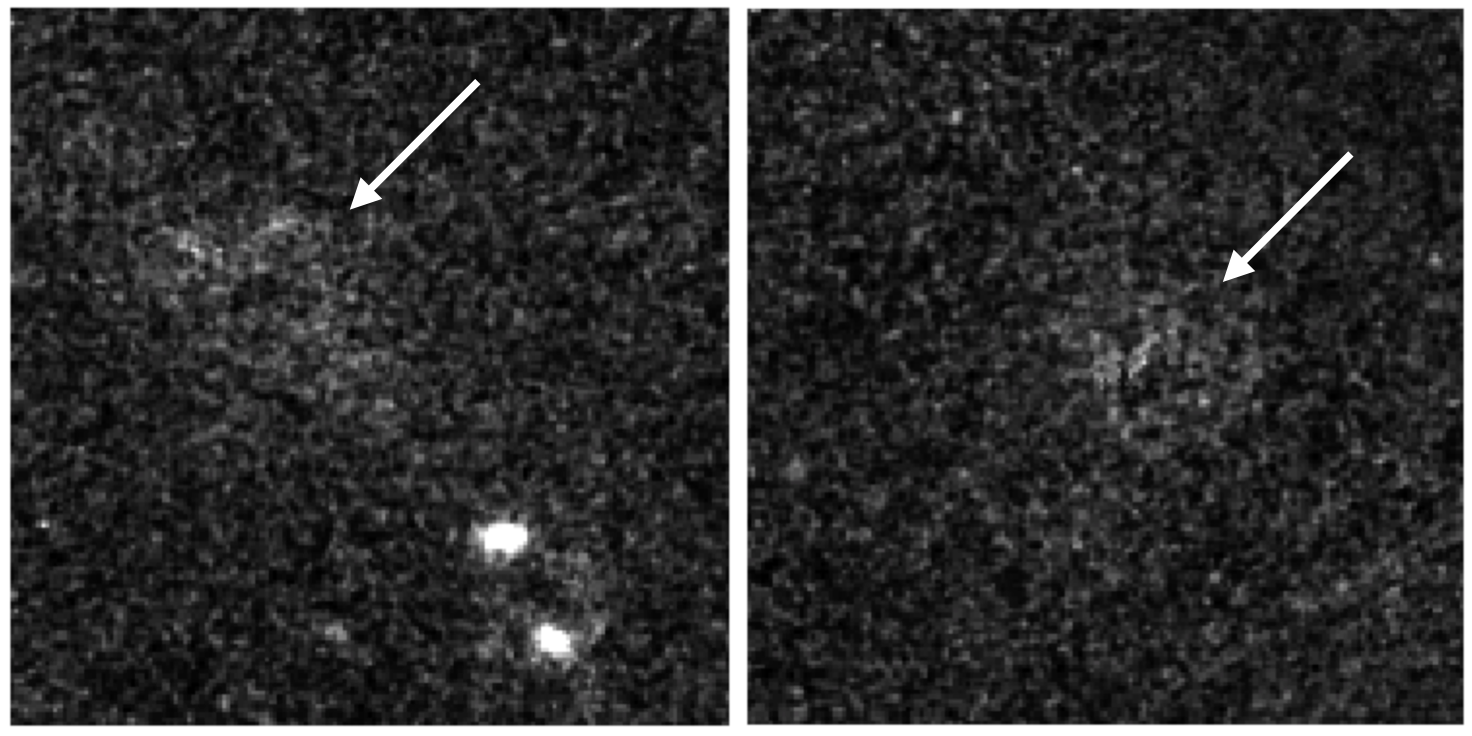}
    \caption{Examples of uncorrected cosmic ray showers in $10\arcsec\times10\arcsec$ cutouts from the final F560W mosaic (see Section~\ref{sec:stage3}).  Approximately ten such artifacts show up across SMILES as $5\sigma$ detections in our photometric catalog (Section~\ref{sec:phot}) after running the correction with CRDS context 1188. }

    \label{fig:crs}
\end{figure}

The correction for cosmic ray showers was implemented in {\tt Pipeline} v1.8.0 as an optional part of the Jump Detection step.  Cosmic ray showers manifest as a wide area, low-level excess in counts below the typical detection threshold for a direct cosmic ray hit.  Unlike the near-infrared detector snowballs, showers in MIRI do not have a saturated core and are not circular, making detection and removal difficult.

We apply the cosmic ray showers correction to the F560W - F1500W observations, finding that the correction removes or improves many, but not all visible shower events.  Source detection (Section~\ref{sec:source_det}) plus visual inspection against NIRCam images finds on order 10 cosmic ray showers are missed that are significant enough to produce a $5\sigma$ artifact picked up by our photometry algorithms, a rate of $\sim0.3$ arcmin$^{2}$ or $\sim0.7$ per MIRI FOV.  Two examples are shown in Figure~\ref{fig:crs}.

We do not apply the correction to F1800W-F2550W as testing showed that low-level arcsec to arcmin scale artifacts were introduced into the science and error images.  This is possibly due to the correction not being optimized in the regime where the noise is dominated by thermal background from the telescope. 
We note that skipping the correction for F1800W-F2550W was later made the default in CRDS context 1202.

The output of stage 1 is uncalibrated count rate  (i.e. the slope or {\tt \_rate}) images, which are fed into stage 2.

\subsection{Stage 2: Image Calibration}\label{sec:stage2}

Stage 2 of the pipeline ({\tt calwebb\_image2} for imaging) produces calibrated science images for each individual exposure.  This includes associating a WCS object with each exposure, dividing out a flatfield, applying flux calibrations, and resampling each exposure into a distortion-corrected 2D image.  We adopt the defaults for this stage in the pipeline minus background subtraction, then apply additional processing to remove warm pixels (Section~\ref{sec:warmpix}), apply custom background subtraction (Section~\ref{sec:background}), and correct the astrometry (Section~\ref{sec:astrometry}) on the resulting calibrated {\tt \_cal} files.

\begin{figure*}[th!]
    \centering
    \includegraphics[width=1.9\columnwidth]{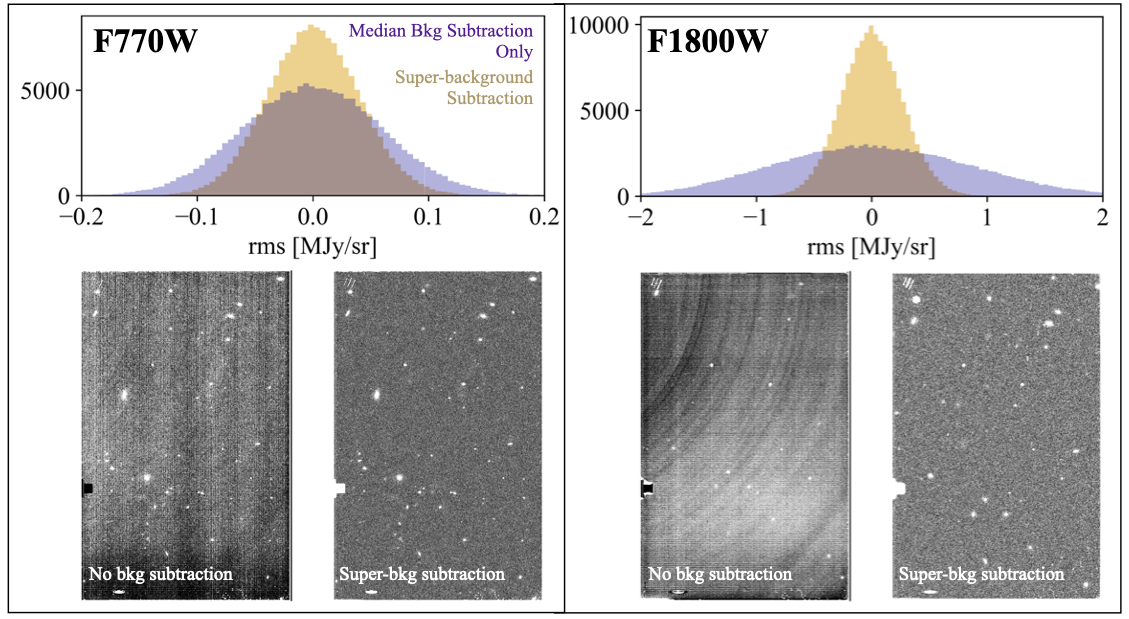}
    \caption{F770W (left) and F1800W (right) {\tt \_cal} exposure from stage 2 before and after super-background subtraction. Note the row/column striping and curved ``tree ring'' features in addition to the background gradient (Section~\ref{sec:background}).  The histograms (top) show that the super-background subtraction significantly reduces the rms over just subtracting the median sky level.}

    \label{fig:bkg_sub}
\end{figure*}

\subsubsection{Removing Warm Pixels}\label{sec:warmpix}

The default pipeline contains two levels of outlier detection, in stage 1 during the Jump Detection Step (Section~\ref{sec:stage1}) and in Stage 3 (Section~\ref{sec:stage3}).  Early generation of SMILES mosaics through stage 3 revealed that faint outliers had escaped both detection steps, resulting in noise peaks in the resampled image with an often distinctive morphology: a central bright pixel with four bright adjacent cross-pixels.  Tracing these noise peaks back to the individual {\tt \_cal} files identified them as warm pixels not masked by the pipeline and not flagged as outliers as they are common across all (contemporary on the order of 1-1.5 months) {\tt \_cal} files in the program. To manually mask these warm pixels, we median stack all {\tt \_cal} files on a per-filter basis and then flag pixels via the DQ mask that are $>3\sigma$ above the median of all pixels of the median-stacked image.  This method is robust in masking real sources as we have 60-66 {\tt \_cal} files over 15 different pointings in a relatively uncrowded field, ensuring that the stacked median image is comprised of background sky noise. The percentage of pixels flagged as warm ranges from 0.03-0.18$\%$ (corresponding to $\sim300-2000$ pixels) across the 8 filters.

\subsubsection{Custom Background Subtraction}\label{sec:background}

\begin{figure}
    \centering
    \includegraphics[width=\columnwidth]{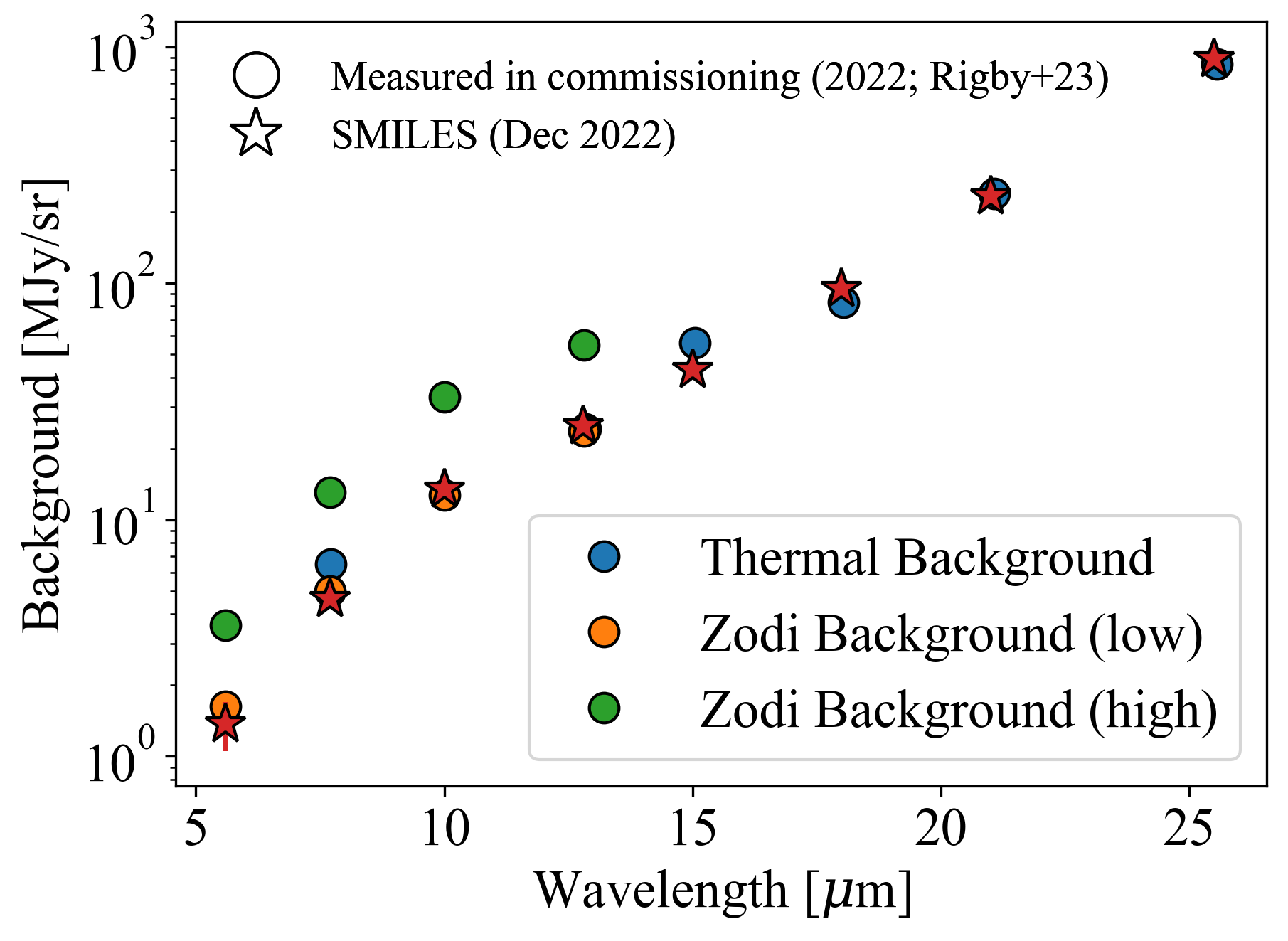}
    \caption{The background level in the SMILES mosaics (stars, observed Dec 2022) compared to low zodi background (orange), high zodi background (green), and thermal background (blue) measured during commissioning in early 2022 \citep{rigby2023a}.}
    \label{fig:background}
\end{figure}

Backgrounds in MIRI are a combination of sky background (dominated by zodi at shorter wavelengths, then thermal emission and scattering from the telescope at $\gtrsim15\,\mu$m \citep[Figure~\ref{fig:background};][]{glasse2015, rigby2023a} and detector effects that can cause large-scale gradients, row/column striping and tree rings that are largely static in detector coordinates for a given dataset \citep{morrison2023}.  As discussed in \citep{dicken2024}, these latter effects are likely residuals from the current flat fielding and/or the dark/reset anomaly correction in stage 1.  Future improvement in these steps will correctly address these multiplicative effects; however, for now, we address them using background subtraction.  This approach has been found to result in a $<5\%$ difference in measured fluxes and is sub-dominant to other flux measurement uncertainties \citep{dicken2024}.

As an extragalactic field with no extended emission, we didn't obtain dedicated background observations for SMILES. In early testing, we opted to manually create a median combined background from the science images on a per-pointing basis, under the assumption that the background may significantly vary spatially, 
using the strategy described in \citet{dicken2024}. We found, as that work did, 
that small dither steps resulted in over-subtraction forming a `halo' around luminous sources (see their Figure 22).  Second, this technique didn't fully address the row/column striping or large-scale noise gradients as described above. 

To improve on this, we instead adopted the ``super-background'' strategy \citep[][see also \cite{yang2023} for a similar procedure]{perez-gonzalez2024}, which takes advantage of our large number of dithers ($60-66$) over 15 contemporaneous (within 1.5 months) pointings.  This strategy works as follows: we start with a full reduction without background subtraction using the standard {\tt Pipeline} to the final Stage 3 mosaics.  From this, we create a source mask with a high ($10\sigma$) threshold using the {\tt photutils} {\tt create\_source\_mask} routine after performing basic background subtraction using {\tt Background2D}.  This source mask is then mapped back onto the individual {\tt \_cal} files in detector space
and median filtering in rows and columns and subtraction of large-scale 2D gradients (again using {\tt Background2D} is performed on the masked {\tt \_cal} files.  These filtered {\tt \_cal} files are then reprocessed through the {\tt Pipeline} to final mosaics.  This process is repeated two more times with successively lower source detection thresholds as well as dilation of the segmentation maps to capture faint, extended emission and extended features of the PSF (see Section~\ref{sec:photometry}).

The final robust source mask is then mapped back to the original, \textit{un}-filtered {\tt \_cal} files and the median of each {\tt \_cal} file is subtracted to bring the median background level to zero.  Then for each unfiltered {\tt \_cal} file, a stacked median `super-background' is created from all other zero-medianed, masked {\tt \_cal} files.  This super-background is scaled back to the median of the {\tt \_cal} file being corrected, and then subtracted.  Finally, row/column filtering and subtraction of any remaining 2D background is done.  These background-subtracted and filtered {\tt \_cal} files are then astrometry-corrected (next section) and fed to Stage 3. 

Figure~\ref{fig:bkg_sub} compares a F770W {\tt \_cal} pre- and post-background subtraction.  The pixel rms is reduced roughly by a factor of two.  Figure~\ref{fig:background} shows that the background levels of our mosaics are in excellent agreement with the zodi and thermal backgrounds measured during commissioning in 2022 \citep{rigby2023a}. A discussion of the final mosaic quality will be presented in Section~\ref{sec:stage3}.

\begin{figure*}[th!]
    \hspace{-40pt}
    \includegraphics[width=2.3\columnwidth]{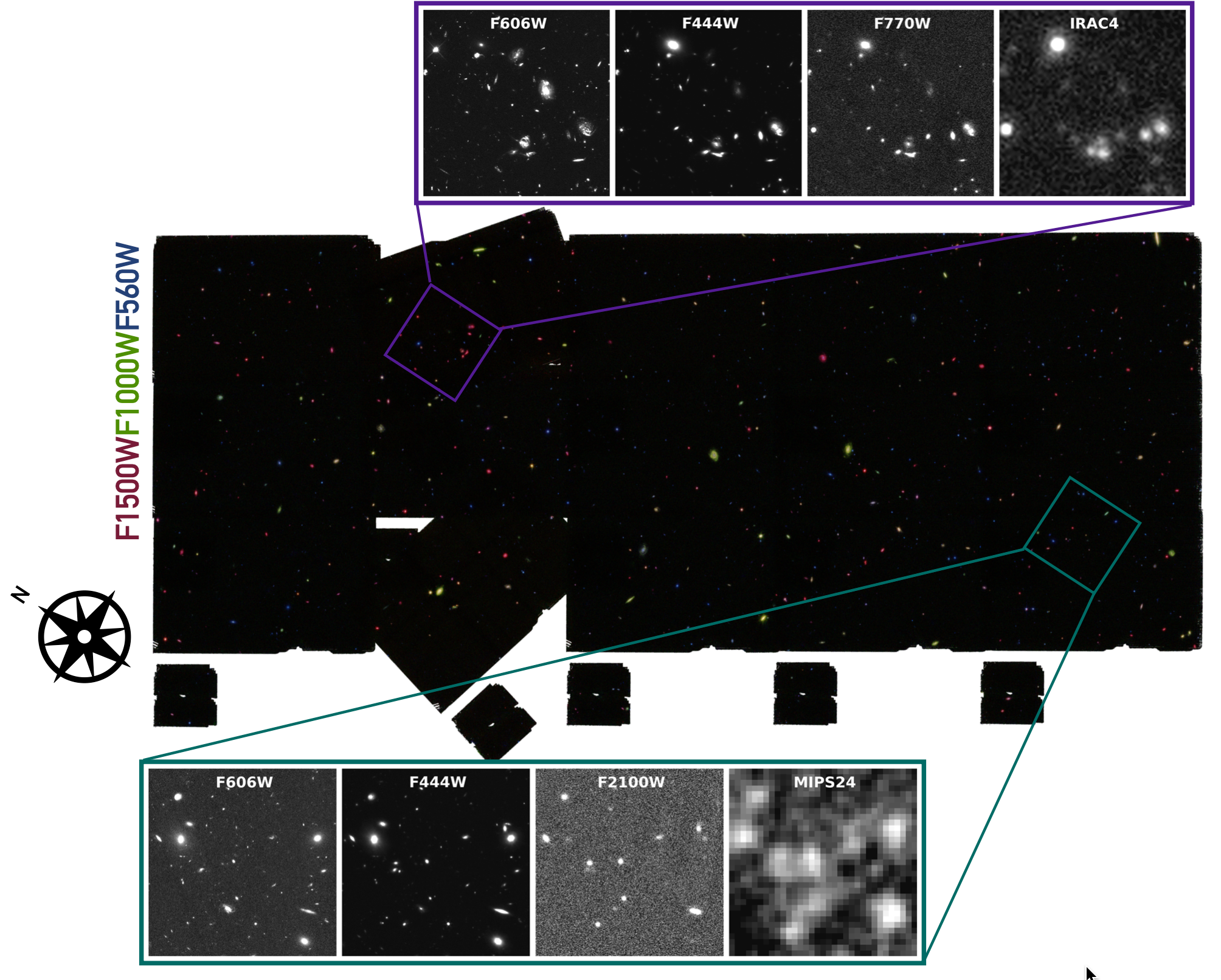}
    \caption{A F1500W, F1000W, F560W RGB image of SMILES.  The zoomed-in areas show $40\arcsec\times40\arcsec$ regions in HST F606W, NIRCam F444W, and matched MIRI and Spitzer bands, [F770W and IRAC Ch 4, top] and matched MIRI and MIPS bands [F2100W and MIPS24, bottom].  MIRI's resolution at $7.7\,\mu$m is only $2\times$ coarser than NIRCam/F444W and $\sim7\times$ finer than IRAC. This gain at $21\mu$m compared to MIPS allows MIRI to be background- rather than confusion-limited across its wavelength range.  Cutout sets are shown at the same stretch. }

    \label{fig:rgb}
\end{figure*}

\subsubsection{Astrometry Correction}\label{sec:astrometry}

\begin{figure*}[th!]
    \centering
    \includegraphics[width=0.65\columnwidth]{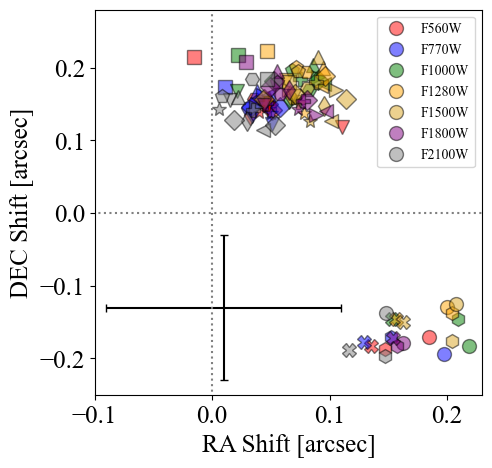}
    \includegraphics[width=0.65\columnwidth]{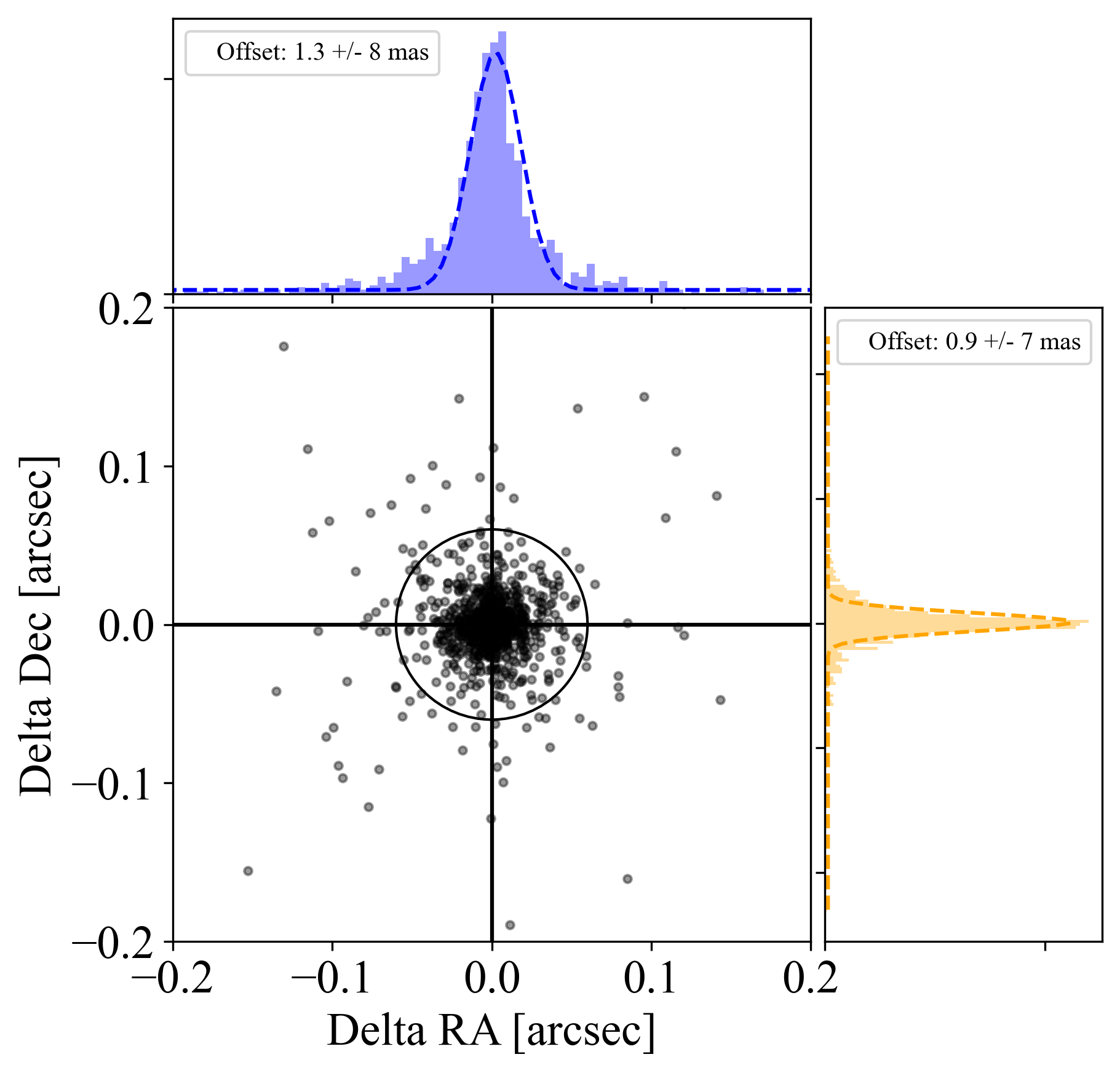}
    \includegraphics[width=0.65\columnwidth]{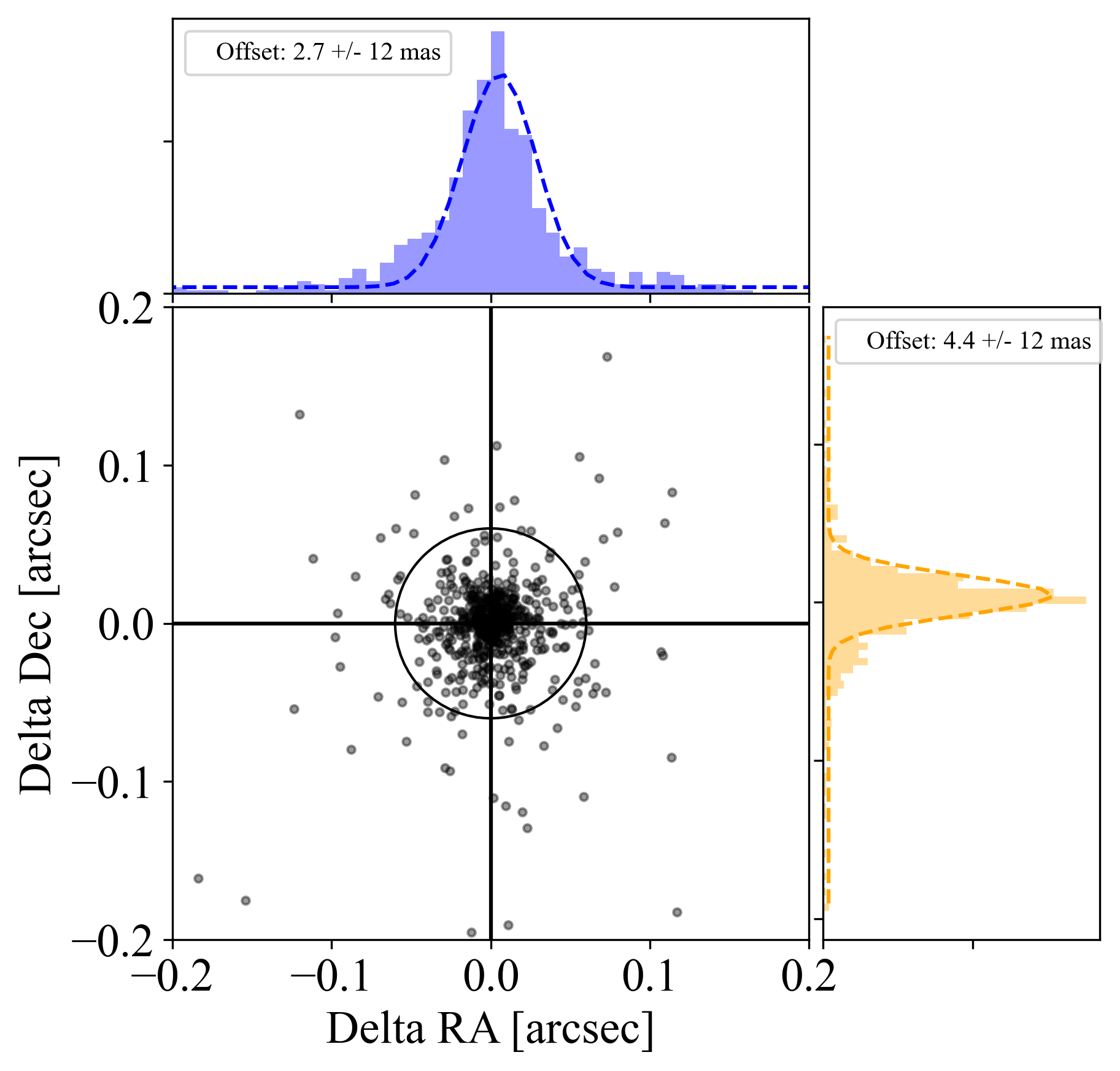}
    \caption{(left) The shifts in RA and Dec applied to each tile (represented by different symbols) and filter (represented by different colors) to correct their astrometry.  The two groups demonstrate that astrometry corrections should be applied on a per-tile basis. The largest shifts are $\sim0.2\pm0.1\arcsec$, within $3\sigma$ of the JWST absolute pointing accuracy ($1\sigma$ is shown by the error bar). (middle, right) The residual offsets between the MIRI corrected catalog and its reference catalog for F770W, using F560W as a reference (middle) and F1500W, using F1280W (right). The astrometric residuals for these two bands are 8-12 mas.}

    \label{fig:astrom}
\end{figure*}

We perform astrometry corrections on the background-subtracted {\tt \_cal} files prior to stage 3.  Due to early limitations in the astrometry correction built into the {\tt Pipeline}, we built external routines based on {\tt TweakReg} that enable more flexibility; namely, we required the ability to use an external reference catalog of NIRCam sources, as the low source density of GAIA sources \citep{gaiacollaboration2022} matched to MIRI detections in MIRI's relatively small FOV ($\sim2.3$ arcmin$^2$) makes it difficult-to-impossible to calculate a robust correction.  We note that later versions of the {\tt Pipeline} include the ability to use user-supplied external catalogs in stage 3.

We register the SMILES astrometry to the JADES NIRCam catalog, which itself is astronomically aligned to GAIA as described in \citet{rieke2023}.  Alignment is done on a per-tile basis, as each tile is a new visit and subject to the JWST absolute pointing accuracy (0.1\arcsec, $1\sigma$, radial), which is driven by the guide star catalog.  Dither positioning uses the Fine Guidance Sensor and is expected to have an accuracy of $\sim2-4$ milli-arcsec (mas) rms per axis with good stability\footnote{https://jwst-docs.stsci.edu/jwst-observatory-characteristics/jwst-pointing-performance};  as such, we apply the same correction to all dithers in a tile.

MIRI is known to have filter-dependent boresight offsets; corrections on order $2-40$ ($7-160$) mas in the column (row) direction are applied in the {\tt Pipeline}. Early tests in applying a single additional astrometry correction via our reference catalog to all filters identified an additional uncorrected offset in the SMILES F2100W images.  To quantify this, we generated un-astrometry-corrected mosaics and measured the relative offsets of sources detected at SNR$>20$, taking the F770W image as the fiducial.   We find that the filter-to-filter offsets (after the boresight correction) are likely negligible given the scatter for all the filters except for a $\sim2\sigma$ offset for F2100W (in declination, $55\pm37$ mas).  We note, however, that the scatter may be inflated due to morphological differences of the sources across the wide wavelength range.
As such, we calculate corrections for each filter separately.

To perform the final astrometry correction, we generate a source catalog for each tile (using windowed centroiding and robust deblending, see Section~\ref{sec:catalog}), which is then compared to the reference catalog. We impose a SNR floor for matches (SN$>20$, except for F2550W which uses SN$>10$ for the source catalog), but forego any constraints on morphology after testing found that having a large number of high SNR matches is as good or better than matching only compact sources. Astrometric corrections are derived based on these matches, allowing for shifts and rotations. Given the offset at F2100W discussed above and to measure the astrometric accuracy of our mosaics as well as possible, we perform the correction on a per-filter basis, using NIRCam F444W detections in the JADES catalog for the initial correction at F560W and then marching the correction up the filters (i.e. using a corrected F560W reference catalog to correct F770W, F770W to F1000W, etc) up to F1500W\footnote{Marching up the bands minimizes potential changes in morphology mentioned above, which can affect centroiding and inflate the astrometric scatter.}.  Due to the decreasing number of matches, the F1500W reference catalog is used to correct F1800W and F2100W. A lack of robust matches between F1500W and F2550W (which has a particularly short exposure time given its high background, see Section~\ref{sec:design}) results in a poor correction; as we didn't find a significant offset between F770W and F2550W, we use the F770W correction directly for F2550W.

Visual inspection is done to check the correction for each tile and the {\tt Tweakreg} tolerance parameter is varied from 0.5-2 to improve problematic tiles. We find that for the two tiles that experienced guide star failures, we needed to correct the astrometry separately for each visit to mitigate different offsets introduced by the telescope absolute pointing.  Figure~\ref{fig:astrom} shows the shifts derived for each tile and filter; a typical shift in declination is $0.2\pm0.1\arcsec$, or two MIRI native pixels.  Our final astrometric accuracy ranges from $0.01-0.02\arcsec$ rms in RA and Dec for F560W-F2100W\footnote{The astrometric accuracy at F2550W is slightly worse, $\sim0.02-0.04\arcsec$ rms in RA, Dec, due to the shallowness of the data.}, 1/10 - 1/5th of a MIRI native pixel (Figure~\ref{fig:astrom}), which we confirm by one final comparison between all filters and the JADES centroids. 

\subsection{Stage 3: Final Mosaics}\label{sec:stage3}

Stage 3 of the pipeline ({\tt calwebb\_image3}) produces the final mosaics from the background-subtracted and astrometry-corrected {\tt \_cal} files.  Here, we skip the {\tt tweakreg} and {\tt skymatch} steps due to our previous corrections.  Outlier detection and resampling are performed with the default setup with the exception of 1) setting the pixel scale to $0.06\arcsec$, twice the JADES mosaicked pixel size\footnote{The exact MIRI pixel size is $0.0599894\arcsec$, twice the NIRCam pixel size which is matched to the effective pixel size of HST HLF mosaics.} \citep{rieke2023}, 2) orienting the mosaics such that North is up ({\tt rotation = 0}), and 3) setting the array output size and reference pixel.  The latter ensures alignment of the pixel grid among all filters.  We note that the {\tt Pipeline} generates an error image that includes Poisson and readout noise, but not correlated pixel noise.  We find that these errors are underestimated by factors of $2-3$ compared to random aperture uncertainties measured from the mosaics directly \citep[see Section~\ref{sec:photometry}, see also][for an alternative method for generating an rms map]{yang2023}.

A final F1500W, F1000W, F560W RGB image is shown in Figure~\ref{fig:rgb}, with zoomed panels showing the vast improvement in sensitivity and resolution in the mid-infrared compared to previous facilities \citep[see e.g. Ji et al., 2024, in prep][for spatially resolved studies with MIRI]{magnelli2023, shen2023}.  The depth of the final mosaics (Table~\ref{tbl:mosaic}) is measured by masking sources and mosaic edges (70 pixels deep) and measuring the median rms in boxes ($d=3\times\mathrm{FWHM}$) placed in a mesh across the mosaics using the {\tt photutils Background2D} routine.  Assuming $65\%$ encircled energy (EE) apertures measured from (semi-)empirical Point Source Functions (PSFs, see Section~\ref{sec:photometry}) and a factor of $1.5\times$ for correlated pixel noise due to resampling to smaller pixels\footnote{This factor was derived by measuring the median standard deviation in 500 randomly placed $d=1\arcsec$ apertures on the native and 0.06$\arcsec$ resolution mosaics (L. Costanin, private communication)}, we reach aperture-corrected $5\sigma$ point source sensitivities of $\sim0.2-17\,\mu$Jy [$25.7 - 20.8$ AB] in F560W$-$F2550W.  This is an improvement of $\sim2\times$ over the predicted ETC values\footnote{ETC v3.0 predictions were measured in the same apertures, assuming a background at the approximate position (HUDF) and time (Dec 15, 2022) of SMILES data acquisition.} at F560W-F1800W  (Table~\ref{tbl:mosaic}), indicating the power of robust background subtraction.    The uniformity of the background level and background rms are discussed in Appendix~\ref{app:a}.

\begin{table*}[]
    \centering
    \begin{tabular}{lcccc}
    \hline
    \hline
        Filter & N & Compl. & Compl.  &  Phot. Accuracy  \\
         & $\geq4\sigma$ & $50\%$ [$\mu$Jy] & $80\%$ [$\mu$Jy] & Median [16,$84\%$] \\
         (1) & (2) & (3) & (4)  & (5) \\
    \hline
    F560W & 2,959 & 0.14 & 0.20 & 1.0 [0.7, 1.4] \\
    F770W & 2,407 & 0.13 & 0.18 & 0.99 [0.6, 1.4] \\
    F1000W & 1,254 & 0.26 & 0.36 & 1.0 [0.7, 1.4] \\
    F1280W & 1,050 & 0.38 & 0.54 & 1.0 [0.6, 1.4] \\
    F1500W & 1,020 & 0.45 & 0.63 & 1.0 [0.6, 1.4] \\
    F1800W & 776 & 1.1 & 1.6 & 1.0 [0.7, 1.3] \\
    F2100W & 756 & 1.5 & 2.2 & 0.99 [0.7, 1.2] \\
    F2550W & 337 & 11 & 15 & 1.0 [0.8, 1.3] \\
    \hline
    \end{tabular}
    \caption{Column 1: MIRI filter. Column 2: the number of $\geq4\sigma$ detections in each filter (Section~\ref{sec:cat_properties}). Columns 3-4: the 50 and $80\%$ completeness limits. Column 5: the median and 16, $84\%$ percentiles of the photometric accuracy simulation at the $5\sigma$ detection limit (Table~\ref{tbl:mosaic}).  See Section~\ref{sec:phot_acc} more for details. }
    \label{tbl:completeness}
\end{table*}

\section{Photometric Catalog}\label{sec:catalog}

\subsection{Source Detection and Photometry}\label{sec:phot}

Source detection and photometry are performed using the JADES photometry pipeline \citep[][Robertson, in prep]{rieke2023, eisenstein2023}, optimized for MIRI imaging.  

\subsubsection{Source Detection}\label{sec:source_det}

For source detection, first a SNR ``detection'' image is constructed as the ratio of the inverse-variance weighted stacks of the F560W and F770W science and error mosaics.  We stack only F560W and F770W to maintain high resolution for source deblending; see Section~\ref{sec:phot_checks} for a discussion of a full stack of all bands. Our source detection method is built on {\tt Astropy} \citep{astropycollaboration2022}, {\tt Photutils} \citep{bradley2023}, and {\tt scikit-image} \citep{vanderwalt2014} routines and based on algorithms previously developed to detect low-surface-brightness features in deep imaging \citep[e.g.][]{borlaff2019}. Source detection is first performed on the detection image with a very low threshold (SNR$\,\geq1.5$).  This initial segmentation map is then refined through a series of custom algorithms that perform iterative filtering and source deblending and reblending \citep[][Robertson, in prep]{rieke2023} to first isolate satellites of bright sources and then repair shredding of bright, resolved galaxies.  Filtering and deblending parameters are fine-tuned by hand for optimal results given the MIRI mosaic properties. The segments of the resulting intermediate segmentation map are then dilated and this is used as a mask for a final pass to identify isolated, faint sources removed by previous filtering.  We found that the high pass filtering required to avoid shredding of bright sources needs to be balanced against an aggressive detection threshold in the final step to recover faint, but visually confirmed sources.  We choose an aggressive source threshold of SNR$\geq2.5$ over 4 pixels, at the cost of a significant number of noise peaks in the final segmentation map.  This is discussed further in Section~\ref{sec:phot_checks}.

\subsubsection{Photometry}\label{sec:photometry}

The segmentation map generated in the last section has its edges masked and then photometry is performed based on {\tt photutils} \citep{bradley2023} in all bands.  Source centroids are determined using the ``windowed'' algorithm from Source Extractor \citep{bertin1996}.  Circular aperture photometry is performed in $r=0.25, 0.3, 0.35, 0.5, 0.6\arcsec$ apertures, ranging from approximately the FWHM at F560W and F770W \citep[$0.21\arcsec$ and $0.27\arcsec$, respectively][]{dicken2024} to approximately the $65\%$ encircled energy radii in the longest bands (Table~\ref{tbl:mosaic}).  We additionally measure $2.5\times$ scaled Kron photometry \citep{kron1980} based on windowed centroiding and using the stacked science F560W+F770W images to define the aperture size, which captures extended low SNR emission better than the original stacked SNR detection image.  Kron apertures with a long axis less than $r=0.25\arcsec$ are replaced with circular apertures with that radius, due to Kron photometry becoming increasingly inaccurate for compact and/or faint sources \citep{graham2005}. We note that the Kron apertures are based on the surface brightness profiles of sources at $\sim5-7\mu$m and do not account for the larger PSF or differences in morphology at longer wavelengths.  We explore this further in Section~\ref{sec:phot_checks}.

The photometric uncertainties are measured by placing apertures (equal to the circular or Kron aperture in question) randomly across the source-masked mosaics to capture all sources of noise including correlated pixel noise \citep{labbe2005, quadri2007, whitaker2011, whitaker2019}.  Though our mosaics are fairly uniform (Appendix~\ref{app:a}), the variation in exposure time is accounted for by binning the random apertures in exposure time percentiles and using a power-law scaling between the rms of the counts and aperture size to determine the sky noise contribution for a given source. We find that this method results in uncertainties $2-3\times$ larger than uncertainties derived from the ERR extension provided by the pipeline.  Finally, aperture corrections are applied to the fluxes and uncertainties based on Point Spread Functions (PSFs) for each band.  For F560W and F770W, we adopt high dynamic range empirical PSFs built from high SNR commissioning data (A. Gaspar, private communication) that capture the ``cruciform'', an extended artifact caused by internal reflection in the MIRI detectors \citep{gaspar2021, libralato2023, dicken2024}.  This artifact is not well modeled in WebbPSF as of version v1.2.1. Extended emission of the calibration source is detected out to the edge of the MIRI imager FOV and so our empirical PSFs can be normalized to one at large radii to produce robust aperture corrections.  
As the cruciform becomes weak-to-negligible beyond $\sim10\,\mu$m, we generate PSFs with WebbPSF for the remaining bands. 
We then account for the effects of mosaicking by injecting the WebbPSFs into blank tiles and running them through the {\tt resample} step, which slightly broadens the PSF core \citep{ji2023}.  The $65\%$ EE radius of our PSFs are listed in Table~\ref{tbl:mosaic}.

\subsection{Quality Checks}\label{sec:phot_checks}

\subsubsection{False Positive Detections}\label{sec:false}

Visual inspection of the catalog is done to verify the robustness of the source deblending, centroiding, and Kron apertures.  The false positive rate of the catalog sources is evaluated initially by matching to the JADES catalog\footnote{\url{https://archive.stsci.edu/hlsps/jades/catalogs/hlsp_jades_jwst_nircam_goods-s-deep_photometry_v2.0_catalog.fits}} within $0.3\arcsec$ to identify candidate spurious detections. The JADES NIRCam imaging at F444W reaches $10\sigma$ depths of 29-30 mag \citep{eisenstein2023}, ensuring that all real sources should be detected, save for rare extreme examples of MIRI-only sources \citep{perez-gonzalez2024a}.  From the total catalog of 5,712 sources, we find 2,101 without JADES counterparts, as anticipated by our aggressive source extraction discussed in the previous section. We discard the 1,809 with SNR$_{\rm F560W}<4$ in a $r=0.5\arcsec$ aperture.  We visually inspect the remaining 292 SNR$_{\rm F560W}\geq4$ detections, finding that 82 are real and missing counterparts due to small gaps in the NIRCam coverage or misidentified JADES counterparts, almost all caused by deblending issues from nearby bright sources.  These are added back into our catalog.  Of the remainder of the SNR $\geq4$ false positives, $\sim75\%$  are mischaracterized extended features of brighter galaxies or noise features at the map edges or on the Lyot coronagraph. This implies a true false positive rate of $\sim1-2\%$ at $>4\sigma$ due to remaining noise artifacts.  

As hinted at in Section~\ref{sec:stage1}, spurious noise peaks (true false positives) are visually identified as associated with unidentified warm/hot pixels (Section~\ref{sec:warmpix}) or uncorrected cosmic ray showers.   
This is confirmed by a test estimating the false positive rate by performing source extraction on inverted mosaics (M. Stone et al. 2024, in prep), which finds $\lesssim10$ noise peaks detected with ``fluxes'' above the $80\%$ completeness limit (roughly the $5\sigma$ detection limit, see next section) across the mosaics.  False positives caused by cosmic ray showers would not show up in this test. On the other hand, the large number of ``detections'' at $<4\sigma$ that lack NIRCam counterparts suggests that the false positive rate rises rapidly with decreasing SNR given our aggressive source extraction. This is likely a combination of cosmic ray showers and true background noise fluctuations. From here on, we only consider a catalog cut to $\geq4\sigma$ detections in an aperture with $r=0.5\arcsec$ in either F560W or F770W (a total of 3,096 sources).  No convincing $\geq4\sigma$ MIRI-only detections at F560W or F770W were found, likely due to the depth of SMILES relative to the much deeper MIRI surveys presented in \citet{perez-gonzalez2024a}.

\subsubsection{Completeness and Photometric Accuracy}\label{sec:phot_acc}

The completeness and photometric accuracy of point sources are assessed as a function of flux density by injecting our MIRI PSFs (Section~\ref{sec:photometry}) into the real mosaics and performing source detection and photometry as described in Section~\ref{sec:phot}.  The source injection is described in more detail in Apprendix~\ref{app:b}. In Table~\ref{tbl:completeness}, we report the resulting 50 and $80\%$ completeness levels for each filter.  Our $80\%$ completeness limits are comparable to our $5\sigma$ detection limits (Table~\ref{tbl:mosaic}), likely due to our aggressive source extraction as discussed in Section~\ref{sec:phot}.  The trade-off is the large false positive rate at low SNR quantified in Section~\ref{sec:false}.  We mitigate this drawback by matching to the much deeper JADES NIRCam data to remove spurious sources and by imposing a SNR cut on the release catalog at $\geq4\sigma$ in F560W or F770W.  

The photometric accuracy (ratio of extracted to input flux density) of artificial sources is shown for F560W and F1500W in Figure~\ref{fig:phot_acc}; all filters are shown in Figure~\ref{fig:phot_acc_all} in Appendix~\ref{app:b}.  The median and 16, 84$\%$ confidence intervals at the $5\sigma$ detection limit are shown in the figures and Table~\ref{tbl:completeness}.  In this case, we impose no criteria for detection, only that there is a counterpart in the extracted catalog to each fake source within $0.2\arcsec$ of its input position that is not blended with a neighbor of equal or brighter flux; this is relevant to our release catalog as we impose no SNR cut on bands longer than F770W. We find no systematic offsets in the recovered flux densities of fake sources.  The $1\sigma$ scatter in the measured-to-input ratios is $1.5-2\times$ higher than what is implied by the $5\sigma$ detection limits measured in similar apertures ($r\sim0.5\arcsec$; Table~\ref{tbl:mosaic}); this reflects factors like variations in the noise and background subtraction (Appendix~\ref{app:a}) and/or variations in the PSFs across the mosaics \citep{libralato2023}, minor flux boosting from neighbors, and stochastic effects.  

\begin{figure}
    \centering
    \includegraphics[width=\columnwidth]{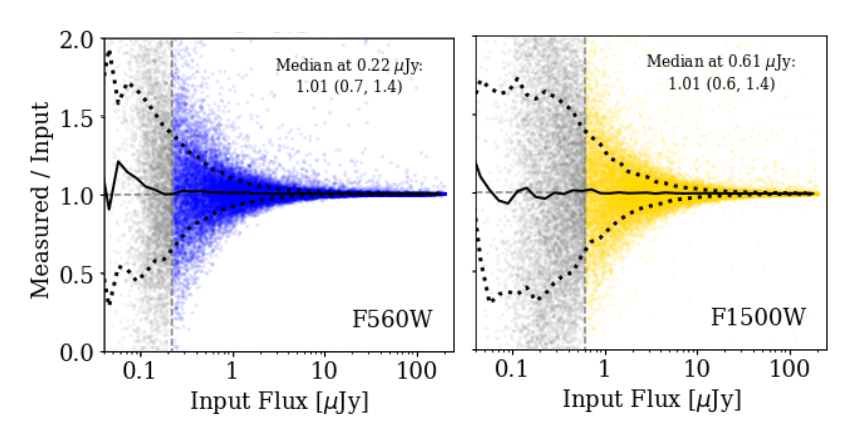}
    \caption{The measured to input flux ratio for fake point sources inserted into the SMILES F560W (left) and F1500W (right) mosaics using a circular aperture with $r=0.5\arcsec$.  The solid (dotted) black lines show the running median ($16, 84\%$ confidence intervals) as a function of input flux. The vertical gray dashed line shows the $5\sigma$ detection limit (Table~\ref{tbl:mosaic}) and the median and confidence intervals at that flux density are listed. }
    \label{fig:phot_acc}
\end{figure}

\subsection{The Effect of using a F560W+F770W Detection Image on Long Wavelength Photometry}

In Section~\ref{sec:phot}, we used a stacked F560W+F770W detection image and perform forced photometry in all bands using the resulting positions and apertures.  To check if this misses sources that are only significantly detected at longer wavelengths, we make a detection image from a stack of all the MIRI mosaics and look for significant ($>4\sigma$) detections in at least 2 bands longer than F770W.  We find 9/35 with NIRCam counterparts; these will be investigated in future analysis. We visually inspect the detections without NIRCam counterparts and conclude they are random noise spikes coincidentally co-spatial in two bands. Looking for single band $4\sigma$ detections in F1000W-F2550W, we find 26 with a NIRCam counterpart, likely strong emission features in faint galaxies.  In general, however, we conclude that we are not missing a significant number of sources by using a F560W+F770W detection image. 

We also examine whether defining the Kron apertures from the F560W+F770W detection image significantly alters the photometry at the longer wavelengths where the PSF is larger. Morphological differences are expected when comparing bands that cover stellar features to those covering dust features. We compare the photometry based on the F560W+F770W detection image to that derived from a stack of all the MIRI filters, finding offsets of $\sim2\%$ ($8\%$) higher fluxes in F1800W-F2100W (F2550W) with $1\sigma$ scatter of $<6\%$ ($<20\%$) at F1800W (F2100W-F2550W) above the $5\sigma$ detection limit.  This indicates that an aperture correction based on individual Kron apertures largely compensates for the larger PSF.

\section{SMILES: NIRSpec Follow-up}\label{sec:nirspec}

\begin{figure*}[th!]
    \centering
    \includegraphics[width=2\columnwidth]{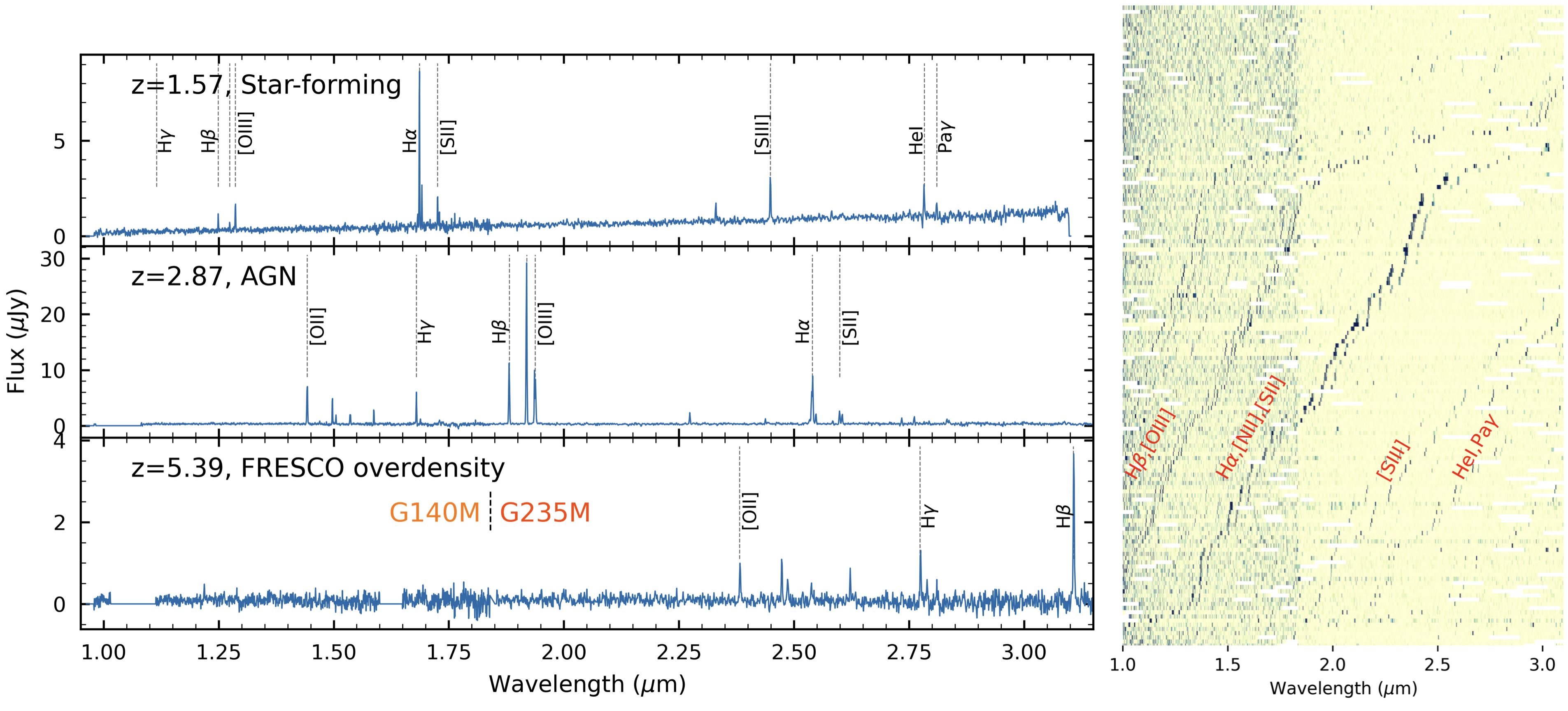}
    \caption{(left) Three example $R\sim1000$ spectra from the SMILES NIRSpec follow-up: a SFG at $z=1.57$ (top), AGN at $z=2.87$ (middle), and a $z=5.39$ [O {\sc iii}] emitter selected from FRESCO \citep{oesch2023} that is part of a galaxy overdensity \citep{helton2024}.  Major emission lines and the boundary between G140M and G235M are labeled.  (right) Stack of 2D spectra for our NIRSpec sample in order of increasing redshift from bottom to top. Areas without data are due to the gap between the NIRSpec detectors.
    }
    \label{fig:ns_spectra}
\end{figure*}

To maximize the science potential of the SMILES MIRI imaging, we included in the program 17 hours of follow-up NIRSpec MOS at $\sim1-3\,\mu$m, which was observed in August 2023. 
The main objective of this follow-up is to provide key information about the ionized gas content of cosmic noon galaxies and AGN. Therefore, the integration times, filter selection, and grating selection were motivated by detecting bright rest-frame optical emission lines at $z\sim 1-3$, where MIRI detects warm dust emission. In this section, we give an overview of the science goals, target selection, and observational setup of this follow-up. The SMILES NIRSpec spectra will be made publically available in a future data release.

\subsection{Science Priorities and Sample}
Optical spectroscopy of SMILES MIRI targets is aimed at measuring properties such as
metallicity, excitation properties of ionized gas, and dust extinction via bright emission lines $-$ [O{\sc ii}]$\lambda 3726, 3729$ ($1.6<z<7.5$);  H$\beta$, [O{\sc iii}]$\lambda4959,5007$ ($1.1\lesssim z\lesssim5.3$); and H$\alpha$, [N{\sc ii}]$\lambda6585$ ($0.5\lesssim z\lesssim3.7$) $-$ primarily at $z\sim 1-2$ where the MIRI bands cover hot dust and multiple PAH features (Figure~\ref{fig:mir}). We additionally cover the extinction-robust strong Paschen lines Pa$\alpha$ at $z\lesssim0.7$ and Pa$\beta$ at $z\lesssim1.45$.
For relatively low mass/low SFR objects and highly dust-obscured systems at $z\sim 1-2$, NIRSpec MOS is uniquely positioned to observe these lines owing to its high sensitivity. The main science priorities of these observations are:

\subsubsection{Star-forming galaxies at cosmic noon}\label{sec:ns_sfg}
   
\noindent\underline{PAH-metallicity relation:} MIRI provides a unique opportunity to robustly characterize the mass fraction of PAH dust grains relative to the total dust mass budget of galaxies \citep{shivaei2024}. Local studies have shown that the PAH fraction is highly dependent on the gas-phase metallicity of galaxies \citep[see ][for a review]{li2020}. On average, this relation has recently been shown through SMILES to hold up to $z\sim2$ \citep{shivaei2024}, albeit using stellar mass as a proxy for metallicity. To expand these results,
We will directly measure metallicity from strong optical lines in a sample of MIRI-detected star-forming galaxies that span a range of mass \citep[and hence, metallicity; e.g.][]{sanders2021}.

\vspace{2mm}
\noindent\underline{Paschen sources:} The footprint of SMILES overlaps significantly with the FRESCO/GOODS-S NIRCam F444W grism observations \citep[Figure~\ref{fig:footprint}][]{oesch2023}, which covers the Pa$\alpha$ (Pa$\beta$) emission line at $z\sim1-1.7$ ($z\sim2-3$). 
Together with our NIRSpec spectroscopy, we can simultaneously observe multiple hydrogen recombination lines (e.g. Pa$\alpha$ and Pa$\beta$ at $z\sim1-1.45$; Pa$\beta$ and H$\alpha$ at $z\sim2-3$) to study the nebular attenuation curve relative to the warm dust emission traced by the MIRI bands.

\vspace{2mm}
\noindent\underline{MIRI-ALMA sources:} Also overlapping SMILES is some of the deepest ALMA data to date both in continuum and spectroscopy \citep[e.g.,][]{walter2016,hatsukade2018,boogaard2023, dunlop2017, adscheid2024}, probing molecular gas, cold dust, and obscured SFR down to the main sequence \citep{popesso2023}. All of the ALMA sources in the primary ASPECS \citep{walter2016} and ASAGAO \citep{hatsukade2018} catalogs have strong detections in multiple bands of MIRI  (Alberts et al., 2024c, in prep) and some have coverage of the Paschen lines from FRESCO. We prioritized targeting ALMA sources with both MIRI and FRESCO coverage to obtain Balmer lines; as above, the combination of Paschen and Balmer lines constrains dust attenuation relative to the warm and cold dust components from MIRI and ALMA.  As a second priority, we targeted known compact ALMA sources using the $\sim0.2\arcsec$ native resolution imaging from ASAGAO. For fillers, we also included ALMA sources with unknown morphologies and/or unconfirmed photometric redshifts. 
The addition of attenuation, metallicity, and ionized gas properties to the existing CO and cold dust continuum emission of ``normal'' galaxies will shed light on the interplay of gas, dust, and metals at cosmic noon.

\vspace{2mm}

\subsubsection{AGN across cosmic time}\label{sec:ns_agn}

As discussed in Section~\ref{sec:agn_science}, \citet{lyu2024} recently identified a large number of obscured AGN in SMILES, including many missed in previous surveys. Spectroscopic follow-up of these AGN candidates can confirm their nature and characterize key AGN properties that are not accessible through the broad-band SED analysis. From the full AGN catalog presented in \citet{lyu2024}, we prioritized obtaining spectra of robust dwarf and high-$z$ AGN candidates as well as ``normal" ($M_*>10^{9.5}~M_\odot$ and $z=0-4$) AGN at cosmic noon.

Depending on the redshift, the SMILES NIRSpec follow-up covers various emission line features relevant to AGN.
For example, for objects at $z\sim1-3.7$, we cover H$\alpha$, H$\beta$, [N{\sc ii}], and [O{\sc iii}] simultaneously, 
enabling us to identify the source of ionizing radiation in galaxies and AGN via rest-frame optical emission line diagnostics like the BPT diagram \citep[e.g.][]{baldwin1981}. For broad-line AGN, we can constrain black hole masses through the line widths of Pa$\alpha$,$\beta$ or He {\sc i} ($z<1.0$), H$\alpha$, and
H$\beta$. 
All these lines can be also
used to search for outflow/inflow signatures that are typically associated with AGN feedback.

\subsubsection{Other Targets}

In addition to prioritized targets discussed above, we added other scientifically interesting targets in the SMILES footprint such as massive (log $\logM\geq10$) quiescent galaxies at $z> 1$ \citep{ji2024}, photometric members of high-redshift ($z\sim5$) overdensities \citep{helton2024}, line emitters at $z> 3$ from the FRESCO grism data \citep{hainline2024}, and radio sources from ultra-deep, high resolution ($\sim0.45\arcsec$) 6 GHz imaging \citep{alberts2020}.

\subsection{NIRSpec Observational Setup}
For the science goals outlined in the previous section, we dedicated three MSA masks with the same settings: we observed in two filter/grating settings G140M/F100LP and G235M/F170LP, providing continuous spectra from $0.97-3.17\,\mu$m with $R\sim 1000$. 
The medium resolution gratings are necessary to deblend closely spaced line pairs like H$\alpha$ and [N{\sc ii}]. For the exposure setup, we observed 16 groups per integration over 2 integrations for each filter, resulting in a total science time of 3.89 hours per mask. We opened 3-shutter slitlets for all the objects and adopted 3-shutter-slitlet nod pattern. 

For the mask design, we adopted the eMPT software \citep{bonaventura2023}. The eMPT is a supplement to the JWST Astronomer's Proposal Tool (APT) that can serve as a replacement for its MSA Planning Tool (MPT) for designing MSA observations. Its advantage is enabling advanced functionality for ambitious science goals, including assigning multiple priorities to different categories of sources such that the maximum number of highest priority targets is observed simultaneously. We gave our most important and/or rare sources the highest priority (e.g., certain AGN and star-forming galaxies with rich ancillary data such as ALMA or FRESCO Paschen lines, 
to ensure a high fraction of slit allocations. The rest of the sample were assigned priorities from 2 to 6, depending on the number count, source concentration, and science driver.
We employed various tests with eMPT and different priority allocations to ensure the most efficient slit allocation strategy among our three masks. 

In the final design, 168 unique sources were successfully allocated slits, with roughly $40\%$ and $20\%$ in the SFG (Section~\ref{sec:ns_sfg}) and AGN (Section~\ref{sec:ns_agn}) categories, respectively.  Figure~\ref{fig:ns_spectra} shows an example spectra for each, as well as a $z>5$ source selected via FRESCO as an [O {\sc iii}] emitter and the stacked 2D spectra of the full sample.

\begin{figure}
    \centering
    \hspace{-10mm}
    \includegraphics[width=1.1\columnwidth]{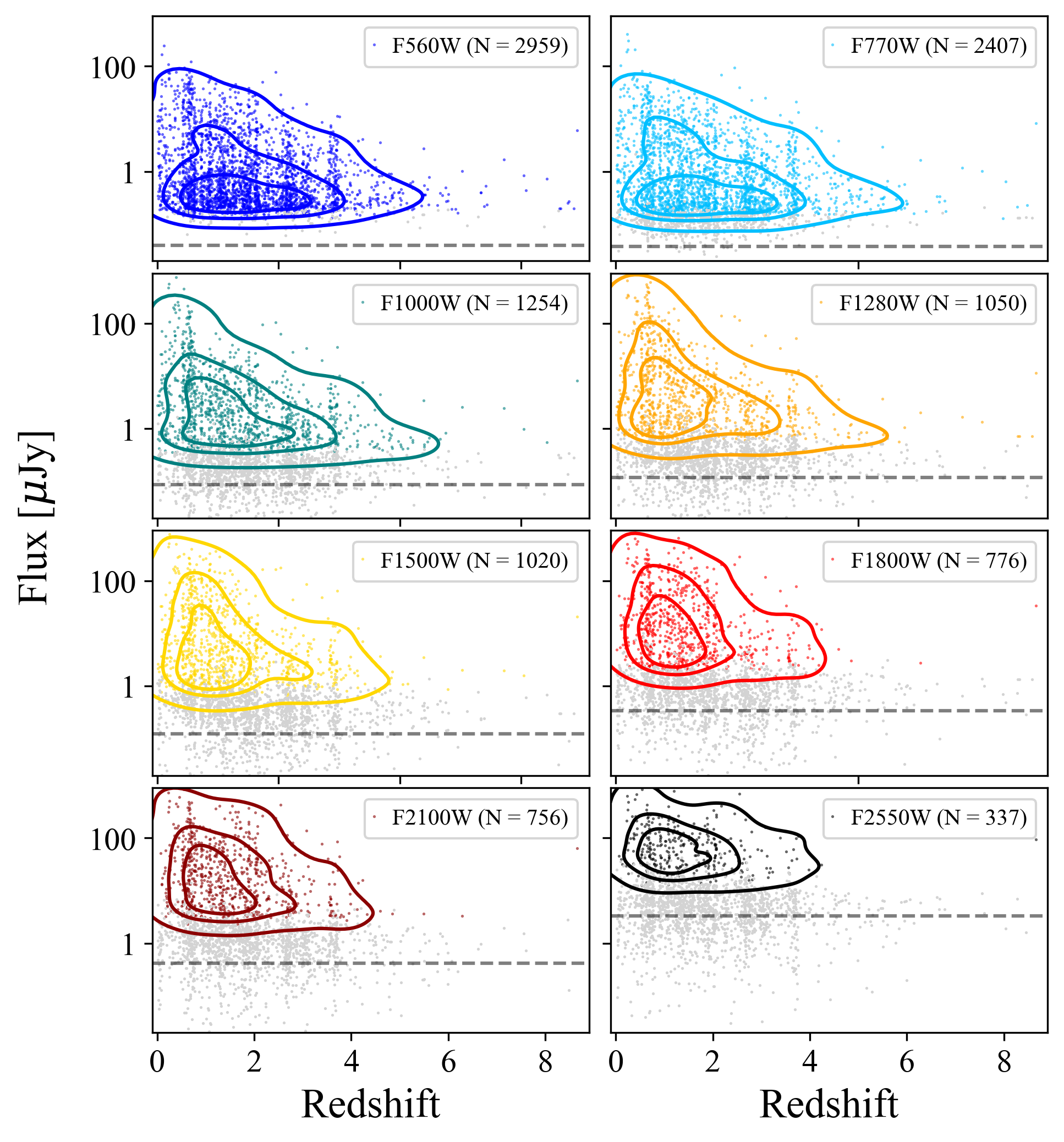}
    \caption{The distribution of the source fluxes as a function of redshift for the SMILES release catalog. Detections at $\geq4\sigma$ are shown in the colored points and contours and lower SNR fluxes are shown in gray.  The number of $\geq4\sigma$ detections is listed in each panel.  The dashed line shows the $1\sigma$ limit. }
    \label{fig:flux_dist}
\end{figure}

\section{SMILES Catalog and Data Release}\label{sec:catalog_and_release}

In this section, we present a brief overview of the demographics of the SMILES photometric catalog and describe the first SMILES data release.  

\subsection{Catalog Properties}\label{sec:cat_properties}
 
In order to quantify the basic demographics of the SMILES photometric release catalog (Section~\ref{sec:release}), the catalog 
is first matched to a compilation of available spectroscopic redshifts in the literature\footnote{The spectroscopic redshift campaigns include (in order of priority for matching), JADES NIRSpec \citep{bunker2023, deugenio2024}, MUSE-Wide \citep{urrutia2019}, MUSE-HUDF \citep{bacon2023}, other surveys based on the CDFS spectroscopic redshift compilation (N. Hathi, private communication), FRESCO \citep{oesch2023}, and ASPECS \citep{walter2016}.} \citep{shivaei2024} and photometric redshifts from JADES \citep{hainline2024}.  1,400 ($45\%$) galaxies in the SMILES photometric catalog have a spectroscopic redshift. Sources without a spectroscopic redshift match are assigned a photometric redshift based on fitting the HST/ACS and NIRCam photometry with the EAZY photo-$z$ fitting code \citep{brammer2008, hainline2024}. Stellar mass estimates are obtained for the MIRI sources from SED modeling (J. Helton, private communication) done using the JADES HST/ACS + NIRCam catalog via the Bayesian fitting code \texttt{Prospector}\footnote{\texttt{Prospector} was run with nest sampler \texttt{dynesty} \citep{speagle2020} and neural net emulator \texttt{parrot} \citep{alsing2020, mathews2023}.} \citep{johnson2021}, following the methodology outlined in \citet{wang2023a, wang2024b} for the \texttt{Prospector-$\beta$} physical model.  Informed observational priors were adopted based on e.g. empirical stellar mass functions, the evolution of the cosmic star-formation rate density, and the stellar mass-stellar metallicity relation. 

\begin{figure}
    \centering
    \includegraphics[width=\columnwidth]{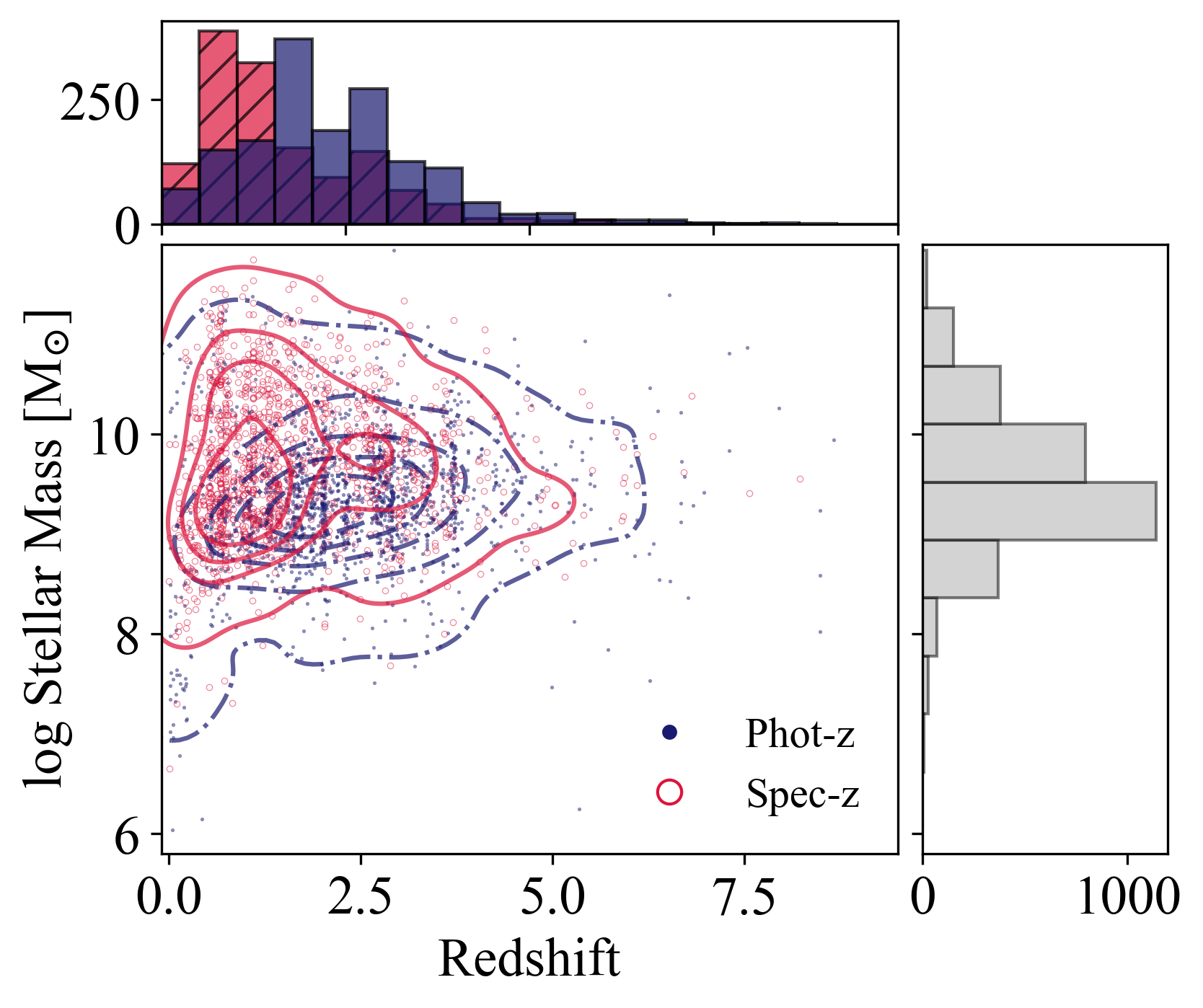}
    \caption{The redshift and stellar mass distribution of the SMILES release catalog.  Spectroscopic (photometric) redshifts are shown in red open circles and solid contours (dark blue points and dot-dashed contours). The top histogram shows the distribution of spectroscopic redshifts (hatched) and photometric redshifts (solid). }
    \label{fig:z_mass}
\end{figure}

The flux distribution and stellar masses of the SMILES catalog sources as a function of redshift are shown in Figures~\ref{fig:flux_dist} and \ref{fig:z_mass}, respectively.  As expected, the distribution of detections is weighted toward low redshift due to the prominent dust features in the MIRI bands up to cosmic noon and the moderate exposure times.  Nevertheless, SMILES detects $\sim90$ galaxies at $z>5$, 21 with confirmed spectroscopic redshifts.  
The SMILES catalog additionally includes $\sim550$ dwarf galaxies with log $\logM<9$.  MIRI number counts in all SMILES bands plus the ultra-deep JADES parallel at 7.7$\,\mu$m (Alberts et al., 2024, in prep) $-$ spanning 4-5 orders of magnitude in flux at $\sim8$ and $\sim24\mu$m when joined with Spitzer surveys $-$ will be presented in Stone et al., 2024, ApJ (submitted).

\subsection{SMILES First Data Release}\label{sec:release}

The first SMILES data release includes eight MIRI mosaics and a photometric catalog with 3,096 sources detected at $\geq4\sigma$ in either F560W or F770W and with forced photometry in all bands, providing limits in the case of non-detections. The data can be accessed at \url{https://archive.stsci.edu/hlsp/smiles}. Data products from the NIRSpec spectroscopic follow-up of the SMILES MIRI imaging (Section~\ref{sec:nirspec}) will be made public in a future release.

\vspace{2mm}
\noindent\underline{Mosaics:} The MIRI mosaics in F560W, F770W, F1000W, F1280W, F1500W, F1800W, F2100W, and F2550W cover $\sim34.5$ arcmin$^2$ in the GOODS-S/HUDF field and reach $5\sigma$ point source sensitivities of $\sim0.2-18$ $\mu$Jy (25.6-20.9 AB; Table~\ref{tbl:mosaic}). The design and reduction of the MIRI mosaics are presented in Sections~\ref{sec:design}-\ref{sec:reduction}.  The data release mosaics each include the following extensions: science image, error image, exposure time map, and weight map.  

\vspace{2mm}
\noindent\underline{Catalog:} The creation and verification of the photometric catalog is described in Section~\ref{sec:catalog}. The release catalog consists of sources detected at $\geq4\sigma$ in either F560W or F770W. As described in Section~\ref{sec:false}, the catalog sources are vetted by matching to the deeper NIRCam catalog and then by visual inspection to recover real sources that are missing NIRCam coverage or where the NIRCam counterpart is mistakenly blended in the JADES catalog.  The flux distribution and number of detections in each band is shown in Figure~\ref{fig:flux_dist} and listed in Table~\ref{tbl:completeness}; the number of $\geq4\sigma$ detections ranges from 2,959 at F560W to 1,254 at F1000W to 756 at F2100W, out of the total of 3,096. Photometry is carried out in apertures defined by a F560W+F770W detection image and a measurement is provided for each source in every band, regardless of significance (with at least F560W or F770W detected at $\geq4\sigma$). Aperture-corrected photometry is provided in 5 circular apertures $r=0.25, 0.3, 0.35, 0.5, 0.6\arcsec$ and in a $2.5\times$-scaled Kron aperture. Photometric uncertainties provided are derived from placing random apertures, as described in Section~\ref{sec:phot}.

\vspace{2mm}
\noindent\underline{Ancillary JWST Data:}
As seen in Figure~\ref{fig:footprint}, SMILES has near-complete NIRCam imaging coverage from JADES \citep{rieke2023, eisenstein2023} in 9 photometric bands as well as partial coverage from JEMS \citep{williams2023} in four NIRCam medium-bands. JADES NIRCam mosaics and catalogs and JADES NIRSpec spectroscopy \citep{bunker2023, deugenio2024} are available at \url{https://archive.stsci.edu/hlsp/jades} and include spectroscopic and photometric redshifts.  JEMS is available at \url{https://archive.stsci.edu/hlsp/jems}.  SMILES additionally has significant overlap with NIRCam grism spectroscopy and imaging from the FRESCO survey \citep{oesch2023}, available at \url{https://archive.stsci.edu/hlsp/fresco}.

\vspace{2mm}
In conclusion, MIRI is emerging as an even more powerful tool in the JWST arsenal than dreamed by many.  We hope that this data release will allow the broader community to explore the full potential of multi-band MIRI surveys at cosmic noon and beyond.


\begin{acknowledgments}
The SMILES team extends their sincerest gratitude to the MIRI instrument team for their tireless work in designing, building, testing, and commissioning MIRI. The authors thank Andras Gaspar for his work constructing empirical MIRI PSFs, Mihai Cara for assistance in building astrometry correction software, and Nimisha Kumari for help in designing and implementing the SMILES NIRSpec observations.  The authors further thank the JADES team for their hard work in creating and publishing the JADES mosaics, spectra, and catalogs. SA, JL, GHR, JM, and MS 
acknowledge support from the JWST Mid-
Infrared Instrument (MIRI) Science Team Lead, grant 80NSSC18K0555, from NASA Goddard Space Flight Center to the University of Arizona. IS acknowledges funding support from the Atracc{\' i}on de Talento program, Grant No. 2022-T1/TIC-20472, of the Comunidad de Madrid, Spain. BER and CNAW acknowledges support from the NIRCam Science Team contract to the University of Arizona, NAS5-02015, and BER additional acknowledges JWST Program 3215. 
\end{acknowledgments}

%



\software{{\tt astropy} \citep{astropycollaboration2022}, 
    {\tt Photutils} \citep{bradley2023}; 
    {\tt scikit-image} \citep{vanderwalt2014}; 
    {\tt eMPT} \citep{bonaventura2024};
    {\tt JWST Calibration Pipeline} \citep{bushouse2023};
    {\tt WebbPSF} \citep{perrin2014}; 
    {\tt Prospector} \citep{johnson2021};
    \texttt{dynesty} \citep{speagle2020};
    \texttt{parrot} \citep{alsing2020, mathews2023}.}



\appendix
\renewcommand\thefigure{\thesection.\arabic{figure}} 

\section{Expanded Mosaic Characteristics}\label{app:a}
\setcounter{figure}{0}

\begin{figure}
    \centering
    \includegraphics[width=0.8\columnwidth]{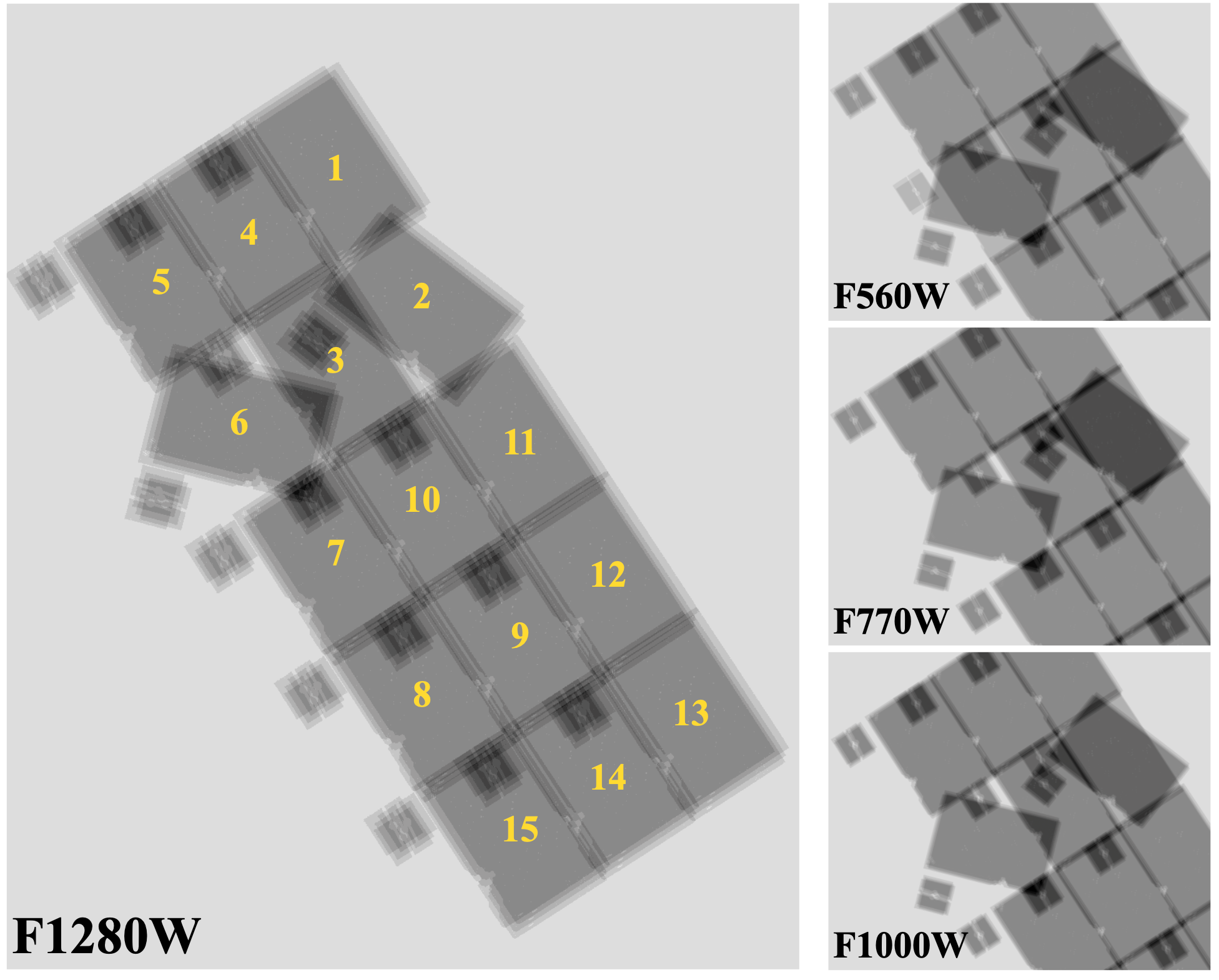}
    \caption{(left) The exposure time map for F1280W, representing all longer bands in terms of spatial distribution of the relative exposure time per pixel.  Total exposure times over 4 dithers can be seen in Table~\ref{tbl:mosaic}.  The exposure time maps are largely uniform, save for where the Lyot chronograph overlaps an adjacent pointing. The tiles are labeled in order of acquisition. Due to guide star failures, Tiles 2 and 6 were re-observed at a later date and different PA, see text for details. (right) Zoom-ins of the exposure time maps for F560W, F770W, and F1000W, which obtained 0-4 exposures before guide star failures in Tiles 2 and 6.   }
    \label{fig:exp_time}
\end{figure}

In this appendix, we give further details about the SMILES MIRI mosaics in filters F560W, F770W, F1000W, F1280W, F1500W, F1800W, F2100W, and F2550W.  The exposure time map for F1280W and pointing order for SMILES can be seen in Figure~\ref{fig:exp_time} (left); the relative exposure time per pixel for F1280W is representative of all of the longer wavelength filters.  The majority of the areas in these mosaics has 4 dithers and a total exposure time as listed in Table~\ref{tbl:mosaic}.  Areas of overlap, such as where the Lyot coronagraph falls interior to the mosaic have double the exposure time.  Tiles 2 and 6 suffered from guide star failures (Section~\ref{sec:design}); prior to failure observations in F560W, F770W, and F1000W (4, 4, and 2 dithers, respectively) were obtained for tile 2.  Only F560W (2 dithers) was obtained for tile 6.  Tiles 2 and 6 were re-observed on Jan 1 and Jan 28, 2023 at aperture $\mathrm{PA}=52$ and $76$ degrees, respectively.  Zoom-ins of the affected area are show for F560W, F770W, and F1000W are shown in the right panels of Figure~\ref{fig:exp_time}.  The relative exposure time per pixel for the rest of the mosaic in these filters is similar to that shown in the left panel, with very minor differences due to the increasing dither pattern size with increasing wavelength. 

\begin{figure*}[th!]
    \centering
    \includegraphics[width=\textwidth]{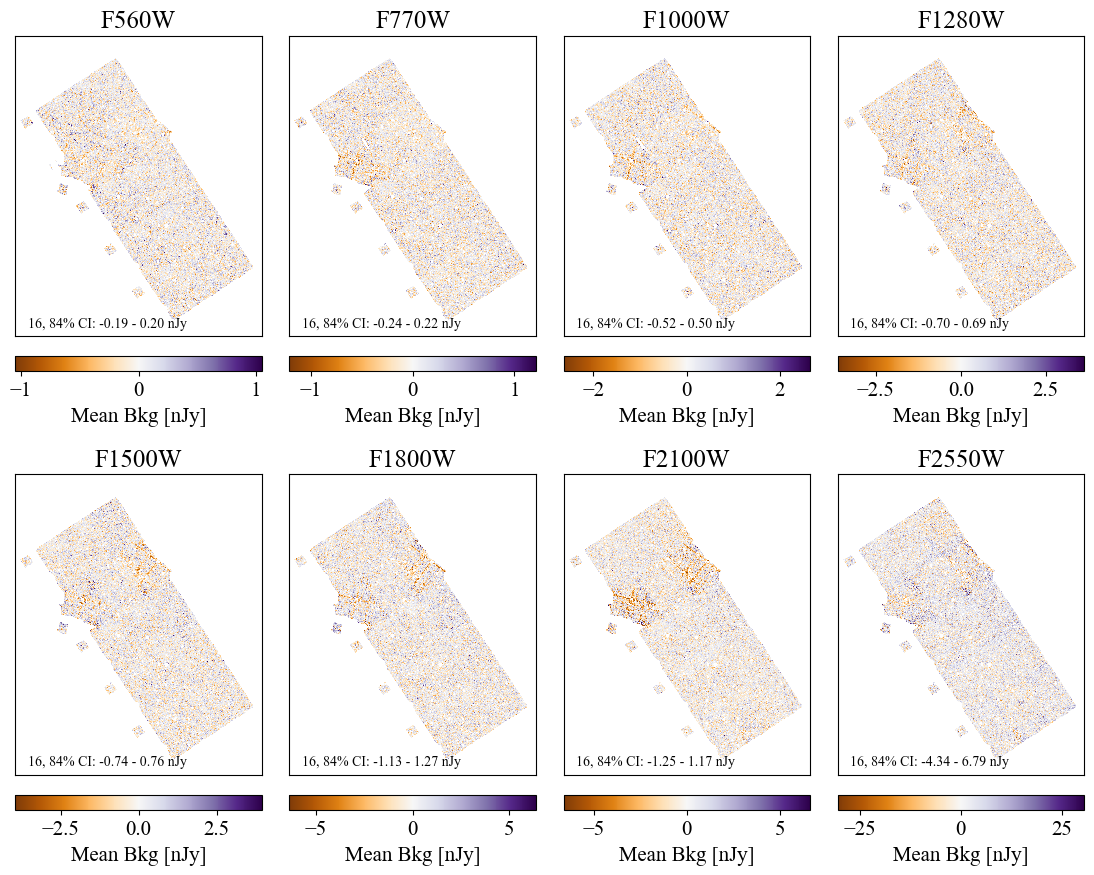}
    \caption{The variation in the sky background of the super-background-subtracted mosaics (Section~\ref{sec:background}) averaged over $65\%$ EE apertures for each filter (see Table~\ref{tbl:mosaic}).  Sources have been masked. These background images are $5\sigma$-clipped to make the variation more visible; the 16 and 84$\%$ confidence intervals (prior to $\sigma$-clipping) are listed. Higher residuals after background subtraction can be seen mainly in Tile 2 and Tile 6 $-$ which were re-observed in Jan 2024 due to guide-star failures (Section~\ref{sec:design}) $-$ at a level of a few to tens of nano-Janskys, well below the typical rms (see Figure~\ref{fig:rms_plots}). }
    \label{fig:bkg_plots}
\end{figure*}

\begin{figure*}[th!]
    \centering
    \includegraphics[width=\textwidth]{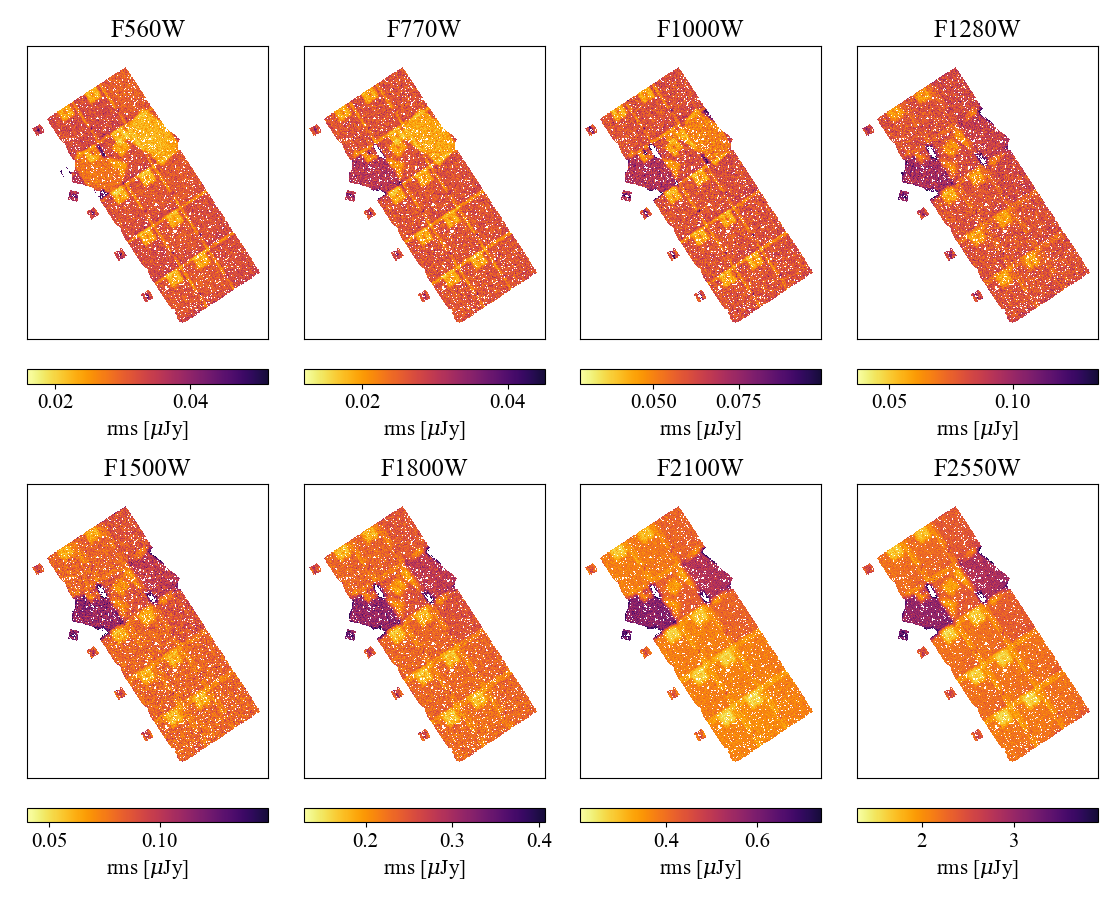}
    \caption{The rms after background subtraction (Section~\ref{sec:background}), averaged over $65\%$ EE apertures (Table~\ref{tbl:mosaic}) for the SMILES mosaics.  No aperture correction has been applied and sources have been masked. As in the residual sky background images (Figure~\ref{fig:bkg_plots}), the rms is highest in Tile 2 and 6, except for Tile 2 F560W - F1000W  and Tile 6 in F560W which have $>4$ dithers as some data was obtained before the guide star failure. }
    \label{fig:rms_plots}
\end{figure*}

In Sections~\ref{sec:background} and \ref{sec:stage3}, we describe our custom background subtraction via the construction of super-backgrounds and the resulting final mosaics.  In Figure~\ref{fig:bkg_plots}, we show the variation in the average background after super-background subtraction, with sources and mosaic edges masked and averaged over apertures matched to the $65\%$ EE (Table~\ref{tbl:mosaic}) for each MIRI filter.  The distribution of sky pixel values after background subtraction is centered on zero (see also Figure~\ref{fig:bkg_sub}) and largely uniform to first order, demonstrating proper subtraction.  Deviations outside the $1\sigma$ confidence intervals can be seen to peak in Tile 2 and 6 (Figure~\ref{fig:exp_time}), where exposures were obtained spread over a 1-1.5 month period after guide star failures.  These deviations are minimal, however, on order of a few to a few tens of nJy.  In Figure~\ref{fig:rms_plots}, the rms is shown for comparison to be on order a few tens to hundreds of nJy, measured in the same size apertures.  Tiles 2 and 6 have higher rms for F770W-F2550W where no exposures were obtained prior to guide star failure.  According to ETC, the background rms from Dec 15, 2022 to Jan 15, 2023 in GOODSS rises by $8-2\%$ from F560W to F1280W, likely due to rising zodi \citep{rigby2023a}. The thermal background levels are expected to be stable over the timescale relevant to these observations \citep[][see also Figure~\ref{fig:background}]{rigby2023a} . Here we are seeing increases of $\sim10-60\%$ in typical rms in tiles 2 and 6, with tile 6 (which was re-observed last, nearly a month after original observations) seeing the largest increase. This suggests a time-dependent factor that is not being wholly addressed by the super-background subtraction.

\section{Photometric Accuracy in SMILES}\label{app:b}
\setcounter{figure}{0}

\begin{figure*}[th!]
    \centering
\includegraphics[width=0.5\columnwidth]{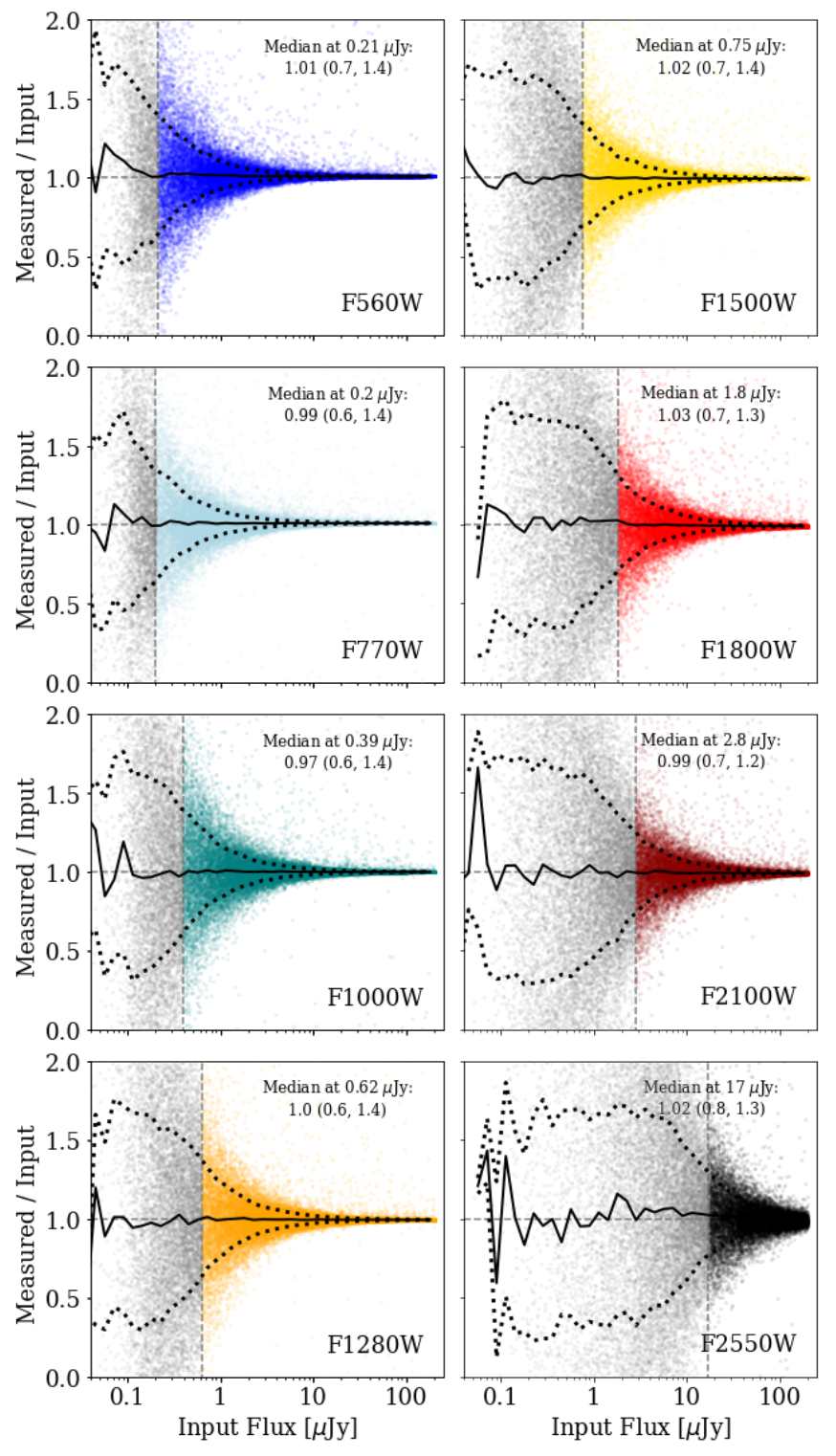}
    \caption{The photometric accuracy represented as the measured flux over the input flux of fake point sources inserted into the F560W-F1280W mosaics (left) and F1500W-F2550W mosaics (right). Flux measurements are using a circular aperture with $r=0.5\arcsec$.}
    \label{fig:phot_acc_all}
\end{figure*}

Completeness and photometric accuracy testing were performed on all eight MIRI mosaics by injecting fake point sources, represented by the PSFs described in Section~\ref{sec:photometry}. The PSFs are scaled to a randomly chosen total flux in 43 flux bins with widths of 0.1 dex ranging from $0.01-200$ $\mu$Jy.  To avoid changing the source density and noise properties of the mosaics, we limit the number of injected PSFs to 150 ($\sim5\%$ of the total source catalog) placed in random positions.  We repeat this for 10 simulations, totalling 1,500 fake sources per filter, per bin. Source detection is run as described in Section~\ref{sec:source_det}. The resulting measured-to-input flux ratios are presented in Figure~\ref{fig:phot_acc_all} for all bands in a $r=0.5\arcsec$ aperture, on average representative of the $65\%$ EE apertures used to measure the rms of the mosaics in Section~\ref{sec:stage3}.  The median and [$16,84]\%$ confidence intervals are shown in each figure for the $5\sigma$ detection limit.

\clearpage

\bibliography{main}{}
\bibliographystyle{aasjournal}



\end{document}